%% file: artiti106v1_main_APS_for_upload.tex
\documentclass[aps,pra,twocolumn,groupedaddress]{revtex4-2}

\usepackage{epsfig}
\usepackage{amssymb}
\input{bss_com}
\input{qip_com}
\input{artiti106v1_eqdef}
\usepackage{color,pstcol}
\newcommand{\ymodifartitionehundredsixvonestepone}[1]
{#1}

\begin{document}

\title{%
Quantum process tomography 
with unknown
single-preparation
input states%
}

\author{%
\ymodifartitionehundredsixvonestepone{Yannick Deville}%
}
\email[]{yannick.deville@irap.omp.eu}
\affiliation{%
\ymodifartitionehundredsixvonestepone{%
{%
Universit\'e de Toulouse}%
,
UPS, CNRS, CNES,
OMP,\\
IRAP
(Institut de Recherche en Astrophysique et Plan\'etologie),
F-31400
Toulouse,
France%
}
}

\author{\ymodifartitionehundredsixvonestepone{Alain Deville}}
\email[]{alain.deville@univ-amu.fr}
\affiliation{%
\ymodifartitionehundredsixvonestepone{%
Aix-Marseille Universit\'e,
CNRS%
,
IM2NP UMR 7334%
,
F-%
13397
Marseille, France%
}
}

\date{\today}

\begin{abstract}
Quantum Process Tomography (QPT) methods aim at
identifying, i.e. estimating,
a given quantum process.
QPT 
is a major quantum information processing 
tool, since it especially allows one to characterize
the actual behavior of
quantum gates,
which are 
the building blocks
of 
quantum computers.
However, usual QPT procedures are complicated,
since they set several constraints on the quantum
states used as inputs of the process to be characterized.
In this paper, we extend QPT so as to avoid two
such constraints.
On the one hand, usual
QPT methods requires one to know, 
hence to precisely control
(i.e. prepare), 
the 
specific
quantum states used as
inputs of the 
considered quantum process,
which is cumbersome.
We therefore propose a Blind, or unsupervised,
extension of QPT 
(i.e. BQPT), 
which means that this approach uses
input 
quantum states whose
values are unknown and arbitrary, except that they are requested to
meet some general known
properties
(and
this approach exploits the output states of
the considered quantum process).
On the other hand, usual QPT methods require 
one
to be able to prepare many copies of the \emph{same} 
(known) input state, 
which is constraining.
On the contrary, we propose ``single-preparation methods'',
i.e. methods which can 
operate with 
only
one
instance 
of each considered input state.
These two new concepts
are here
illustrated 
with practical BQPT methods which are numerically validated,
in the case when: 
\ymodifartitionehundredsixvonestepone{i) random 
pure
states are used as
inputs and their
required properties are especially related to the
statistical independence 
of the random variables that
define them,
ii) the considered quantum process is
based on cylindrical-symmetry Heisenberg 
spin
coupling.}
These concepts
may 
be extended 
to a much
wider class of processes and to 
BQPT methods
based on
other input quantum state
properties.
\end{abstract}

\maketitle

\input{artiti106v1_core_for_upload}

\appendix

\input{artiti106v1_appendix}

\input{artiti106v1_bib}
\end{document}

%% file: bss_com.tex
\newcommand{\Ysssrcnb}
           {M}

\newcommand{\Yssbothtimevalnb}
           {N}
\newcommand{\ysssrcsigdisctimecentvec}
          {$\Ysssrcsigdisctimecentvec$}
\newcommand{\Ysssrcsigdisctimecentvec}
           {s}

\newcommand{\ysssrcsigdisctimecentone}
          {$\Ysssrcsigdisctimecentone$}
\newcommand{\Ysssrcsigdisctimecentone}
           {\Ysssrcsigdisctimecentvec _{1}}

\newcommand{\ysssrcsigdisctimecentlast}
          {$\Ysssrcsigdisctimecentlast$}
\newcommand{\Ysssrcsigdisctimecentlast}
           {\Ysssrcsigdisctimecentvec _{\Ysssrcnb}}

\newcommand{\Yssextquadsrcsigdisctimecentmatrixestim}
{D}
\newcommand{\yssmixsigdisctimecentvec}
          {$\Yssmixsigdisctimecentvec$}
\newcommand{\Yssmixsigdisctimecentvec}
           {x}
\newcommand{\yssoutsepsystsigdisctimecentvec}
          {$\Yssoutsepsystsigdisctimecentvec$}
\newcommand{\Yssoutsepsystsigdisctimecentvec}
           {y}

\newcommand{\yssoutsepsystsigdisctimecentone}
          {$\Yssoutsepsystsigdisctimecentone$}
\newcommand{\Yssoutsepsystsigdisctimecentone}
           {\Yssoutsepsystsigdisctimecentvec _{1}}

\newcommand{\Yssmixmatrixscalarquad}
{B}

\newcommand{\Yssmixmatrixscalarelquadnoindex}
          {b}

\newcommand{\Yssmixmatrixscalarelquadnoindexestim}
          {\hat{\Yssmixmatrixscalarelquadnoindex}}

\newcommand{\Yssmixmatrixscalarquadextestim}
{C}

\newcommand{\Ysssepsystmatrixscalar}
{C}

\newcommand{\Ysssepsystmatrixscalarelnoindex}
           {c}

%% file: qip_com.tex
\newcommand{\Yqubitonesevspace}
{{\cal E}}
\newcommand{\yqubitoneindexstd}
{$\Yqubitoneindexstd$}
\newcommand{\Yqubitoneindexstd}
{i}
\newcommand{\yqubitonespaceindexstd}
{$\Yqubitonespaceindexstd$}
\newcommand{\Yqubitonespaceindexstd}
{\Yqubitonesevspace_{\Yqubitoneindexstd}}
\newcommand{\yqubitonetimeinit}
{$\Yqubitonetimeinit$}
\newcommand{\Yqubitonetimeinit}
{t_0}
\newcommand{\yqubitonetimefinal}
{$\Yqubitonetimefinal$}
\newcommand{\Yqubitonetimefinal}
{t}
\newcommand{\Yqubitonetimeinitstateindexone}
{| \psi_1 ( \Yqubitonetimeinit ) \rangle}
\newcommand{\Yqubitonetimeinitstateindextwo}
{| \psi_2 ( \Yqubitonetimeinit ) \rangle}
\newcommand{\yqubitonetimenonestateindexqubitoneindexstd}
{$\Yqubitonetimenonestateindexqubitoneindexstd$}
\newcommand{\Yqubitonetimenonestateindexqubitoneindexstd}
{| \psi_{\Yqubitoneindexstd} \rangle}
\newcommand{\yqubitbothspace}
{$\Yqubitbothspace$}
\newcommand{\Yqubitbothspace}
{\Yqubitonesevspace}
\newcommand{\ymixsyststateinitial}
{$\Ymixsyststateinitial$}
\newcommand{\Ymixsyststateinitial}
{| \psi 
( \Yqubitonetimeinit ) \rangle
}
\newcommand{\Ytwoqubitwritereadtimeinterval}
{\tau}
\newcommand{\ytwoqubitwritereadtimeintervalindexone}
{$\Ytwoqubitwritereadtimeintervalindexone$}
\newcommand{\Ytwoqubitwritereadtimeintervalindexone}
{\Ytwoqubitwritereadtimeinterval_{1}}
\newcommand{\ytwoqubitwritereadtimeintervalindextwo}
{$\Ytwoqubitwritereadtimeintervalindextwo$}
\newcommand{\Ytwoqubitwritereadtimeintervalindextwo}
{\Ytwoqubitwritereadtimeinterval_{2}}
\newcommand{\ytwoqubitwritereadtimeintervalindexthree}
{$\Ytwoqubitwritereadtimeintervalindexthree$}
\newcommand{\Ytwoqubitwritereadtimeintervalindexthree}
{\Ytwoqubitwritereadtimeinterval_{3}}

\newcommand{\Yqubitbothtimefinalstatecoefindexindexstd}
{j}
\newcommand{\yveccompsyststatetinit}
{$\Yveccompsyststatetinit$}
\newcommand{\Yveccompsyststatetinit}
{C_{+
} 
(
\Yqubitonetimeinit
)
}
\newcommand{\yveccompsyststatetfinal}
{$\Yveccompsyststatetfinal$}
\newcommand{\Yveccompsyststatetfinal}
{C_{+
} (t)}
\newcommand{\ymixsyststatefinal}
{$\Ymixsyststatefinal$}
\newcommand{\Ymixsyststatefinal}
{| \psi (t) 
\rangle
}
\newcommand{\yopmix}
{$\Yopmix$}
\newcommand{\Yopmix}
{M}
\newcommand{\yopmixestim}
{$\Yopmixestim$}
\newcommand{\Yopmixestim}
{\widehat{\Yopmix}}
\newcommand{\Yopmixbases}
{Q}
\newcommand{\Yopmixbasesimagin}
{\Yopmixbases_{\Ysqrtminusone}}
\newcommand{\yopmixdiag}
{$\Yopmixdiag$}
\newcommand{\Yopmixdiag}
{D}
\newcommand{\yopmixdiagestim}
{$\Yopmixdiagestim$}
\newcommand{\Yopmixdiagestim}
{\widehat{\Yopmixdiag}}

\newcommand{\Ysepsystdedicstateoutcoefnot}
{c}
\newcommand{\Ysepsystdedicstateoutcoefpart}
{\Ysepsystdedicstateoutcoefnot_{5}}

\newcommand{\Ysepsystdedicstateoutcoefprobnot}
{P}

\newcommand{\Ysepsystdedicstateoutcoefprobnotox}
{P}

\newcommand{\Ysepsystdedicopmixbasesonestateoutcoefprobzznot}
{\Ysepsystdedicstateoutcoefprobnot}

\newcommand{\Ysepsystdedicopmixbasesimaginonestateoutcoefprobzznot}
{\Ysepsystdedicstateoutcoefprobnot}
\newcommand{\ysepsystdedicopmixbasesimaginonestateoutcoefprobzzplusplus}
{$\Ysepsystdedicopmixbasesimaginonestateoutcoefprobzzplusplus$}
\newcommand{\Ysepsystdedicopmixbasesimaginonestateoutcoefprobzzplusplus}
{\Ysepsystdedicopmixbasesimaginonestateoutcoefprobzznot
_{1z}
( \Yopmixbasesimagin )
}
\newcommand{\ytwoqubitsbasisplusplus}
{$\Ytwoqubitsbasisplusplus$}
\newcommand{\Ytwoqubitsbasisplusplus}
{{\cal B} _+}
\newcommand{\yqubitnbarb}
{$\Yqubitnbarb$}
\newcommand{\Yqubitnbarb}
{Q}
\newcommand{\Yqubitnbarbtimenonestatenot}
{\psi}
\newcommand{\yqubitnbarbtimenonestate}
{$\Yqubitnbarbtimenonestate$}
\newcommand{\Yqubitnbarbtimenonestate}
{| \Yqubitnbarbtimenonestatenot \rangle}
\newcommand{\yqubitnbarbtimenonestateseqindex}
{$\Yqubitnbarbtimenonestateseqindex$}
\newcommand{\Yqubitnbarbtimenonestateseqindex}
{| \Yqubitnbarbtimenonestatenot
(
\Ytwoqubitseqindex
)
\rangle}
\newcommand{\yqubitnbarbstatecoefindexequalstd}
{$\Yqubitnbarbstatecoefindexequalstd$}
\newcommand{\Yqubitnbarbstatecoefindexequalstd}
{j}
\newcommand{\yqubitnbarbspaceoverallvecbasiscoefindexequalstd}
{$\Yqubitnbarbspaceoverallvecbasiscoefindexequalstd$}
\newcommand{\Yqubitnbarbspaceoverallvecbasiscoefindexequalstd}
{
|
{\Yqubitnbarbstatecoefindexequalstd}
\rangle
}
\newcommand{\yqubitnbarbspaceoverallcoefindexequalstdwithvecbasis}
{$\Yqubitnbarbspaceoverallcoefindexequalstdwithvecbasis$}
\newcommand{\Yqubitnbarbspaceoverallcoefindexequalstdwithvecbasis}
{
c
_
{\Yqubitnbarbstatecoefindexequalstd}
}
\newcommand{\yqubitnbarbspaceoverallcoefindexequalstdwithvecbasisseqindex}
{$\Yqubitnbarbspaceoverallcoefindexequalstdwithvecbasisseqindex$}
\newcommand{\Yqubitnbarbspaceoverallcoefindexequalstdwithvecbasisseqindex}
{
c
_
{\Yqubitnbarbstatecoefindexequalstd}
(
\Ytwoqubitseqindex
)
}
\newcommand{\yqubitnbarbspaceoveralleventindexequalstdwithvecbasis}
{$\Yqubitnbarbspaceoveralleventindexequalstdwithvecbasis$}
\newcommand{\Yqubitnbarbspaceoveralleventindexequalstdwithvecbasis}
{
A
_
{\Yqubitnbarbstatecoefindexequalstd}
}
\newcommand{\yqubitnbarbspaceoveralleventindexequalstdwithvecbasisprob}
{$\Yqubitnbarbspaceoveralleventindexequalstdwithvecbasisprob$}
\newcommand{\Yqubitnbarbspaceoveralleventindexequalstdwithvecbasisprob}
{
P
(
\Yqubitnbarbspaceoveralleventindexequalstdwithvecbasis
)
}
\newcommand{\yqubitnbarbspaceoveralleventindexequalstdwithvecbasisprobexpect}
{$\Yqubitnbarbspaceoveralleventindexequalstdwithvecbasisprobexpect$}
\newcommand{\Yqubitnbarbspaceoveralleventindexequalstdwithvecbasisprobexpect}
{
E
\{
\Yqubitnbarbspaceoveralleventindexequalstdwithvecbasisprob
\}
}
\newcommand{\Yqubitnbarbspaceoveralleventindexequalstdwithvecbasisprobexpectapproxonetwoqubitseqnb}
{
E
^{\prime}
\{
\Yqubitnbarbspaceoveralleventindexequalstdwithvecbasisprob
\}
}
\newcommand{\yqubitnbarbspaceoveralleventindexequalstdwithvecbasisprobexpectapproxtwotwoqubitseqnb}
{$\Yqubitnbarbspaceoveralleventindexequalstdwithvecbasisprobexpectapproxtwotwoqubitseqnb$}
\newcommand{\Yqubitnbarbspaceoveralleventindexequalstdwithvecbasisprobexpectapproxtwotwoqubitseqnb}
{
E
^{\prime \prime}
\{
\Yqubitnbarbspaceoveralleventindexequalstdwithvecbasisprob
\}
}
\newcommand{\yqubitnbarbspaceoveralleventindexequalstdwithvecbasisprobseqindex}
{$\Yqubitnbarbspaceoveralleventindexequalstdwithvecbasisprobseqindex$}
\newcommand{\Yqubitnbarbspaceoveralleventindexequalstdwithvecbasisprobseqindex}
{
P
(
\Yqubitnbarbspaceoveralleventindexequalstdwithvecbasis
,
\Ytwoqubitseqindex
)
}
\newcommand{\Yqubitnbarbspaceoveralleventindexequalstdwithvecbasisprobseqindexapproxwritereadonestatenb}
{
P
^{\prime}
(
\Yqubitnbarbspaceoveralleventindexequalstdwithvecbasis
,
\Ytwoqubitseqindex
,
\Ywritereadonestatenb
)
}
\newcommand{\Yfuncnumberofofeventoccur}
{{\cal N}}
\newcommand{\yqubitnbarbspaceoveralleventindexequalstdwithvecbasisnboccurseqindexwritereadonestatenb}
{$\Yqubitnbarbspaceoveralleventindexequalstdwithvecbasisnboccurseqindexwritereadonestatenb$}
\newcommand{\Yqubitnbarbspaceoveralleventindexequalstdwithvecbasisnboccurseqindexwritereadonestatenb}
{
\Yfuncnumberofofeventoccur
(
\Yqubitnbarbspaceoveralleventindexequalstdwithvecbasis
,
\Ytwoqubitseqindex
,
\Ywritereadonestatenb
)
}
\newcommand{\yqubitnbarbspaceoveralleventindexequalstdwithvecbasisnboccurqubitnbarbstatewritereadseqnb}
{$\Yqubitnbarbspaceoveralleventindexequalstdwithvecbasisnboccurqubitnbarbstatewritereadseqnb$}
\newcommand{\Yqubitnbarbspaceoveralleventindexequalstdwithvecbasisnboccurqubitnbarbstatewritereadseqnb}
{
\Yfuncnumberofofeventoccur
(
\Yqubitnbarbspaceoveralleventindexequalstdwithvecbasis
,
\Yqubitnbarbstatewritereadseqnb
)
}
\newcommand{\yqubitnbarbspaceoveralleventindexequalstdwithvecbasisindicfuncqubitnbarbstatewritereadseqindex}
{$\Yqubitnbarbspaceoveralleventindexequalstdwithvecbasisindicfuncqubitnbarbstatewritereadseqindex$}
\newcommand{\Yqubitnbarbspaceoveralleventindexequalstdwithvecbasisindicfuncqubitnbarbstatewritereadseqindex}
{
1\hspace{-.12cm}1
(
\Yqubitnbarbspaceoveralleventindexequalstdwithvecbasis
,
\Yqubitnbarbstatewritereadseqindex
)
}
\newcommand{\ytwoqubitseqindex}
{$\Ytwoqubitseqindex$}
\newcommand{\Ytwoqubitseqindex}
{n}
\newcommand{\ytwoqubitseqnb}
{$\Ytwoqubitseqnb$}
\newcommand{\Ytwoqubitseqnb}
{N}
\newcommand{\yqubitnbarbstatewritereadseqindex}
{$\Yqubitnbarbstatewritereadseqindex$}
\newcommand{\Yqubitnbarbstatewritereadseqindex}
{\ell}
\newcommand{\yqubitnbarbstatewritereadseqnb}
{$\Yqubitnbarbstatewritereadseqnb$}
\newcommand{\Yqubitnbarbstatewritereadseqnb}
{L}
\newcommand{\ytwoqubitsprobaindexstd}
{$\Ytwoqubitsprobaindexstd$}
\newcommand{\Ytwoqubitsprobaindexstd}
{\Ytwoqubitsprobanot
_{\Yqubitbothtimefinalstatecoefindexindexstd}
}
\newcommand{\Ytwoqubitsprobanot}
{p}
\newcommand{\ytwoqubitsprobaplusplusdirzz}
{$\Ytwoqubitsprobaplusplusdirzz$}
\newcommand{\Ytwoqubitsprobaplusplusdirzz}
{\Ytwoqubitsprobanot
_{1zz}
}
\newcommand{\ytwoqubitsprobaplusminusdirzz}
{$\Ytwoqubitsprobaplusminusdirzz$}
\newcommand{\Ytwoqubitsprobaplusminusdirzz}
{\Ytwoqubitsprobanot
_{2zz}
}
\newcommand{\ytwoqubitsprobaminusplusdirzz}
{$\Ytwoqubitsprobaminusplusdirzz$}
\newcommand{\Ytwoqubitsprobaminusplusdirzz}
{\Ytwoqubitsprobanot
_{3zz}
}
\newcommand{\ytwoqubitsprobaminusminusdirzz}
{$\Ytwoqubitsprobaminusminusdirzz$}
\newcommand{\Ytwoqubitsprobaminusminusdirzz}
{\Ytwoqubitsprobanot
_{4zz}
}
\newcommand{\ytwoqubitsprobaindexstddirzz}
{$\Ytwoqubitsprobaindexstddirzz$}
\newcommand{\Ytwoqubitsprobaindexstddirzz}
{\Ytwoqubitsprobanot
_{\Yqubitbothtimefinalstatecoefindexindexstd zz}
}

\newcommand{\ytwoqubitsprobaplusplusdirxx}
{$\Ytwoqubitsprobaplusplusdirxx$}
\newcommand{\Ytwoqubitsprobaplusplusdirxx}
{\Ytwoqubitsprobanot 
_{1xx}
}
\newcommand{\ytwoqubitsprobaminusminusdirxx}
{$\Ytwoqubitsprobaminusminusdirxx$}
\newcommand{\Ytwoqubitsprobaminusminusdirxx}
{\Ytwoqubitsprobanot
_{4xx}
}
\newcommand{\ytwoqubitsprobaindexstddirxx}
{$\Ytwoqubitsprobaindexstddirxx$}
\newcommand{\Ytwoqubitsprobaindexstddirxx}
{\Ytwoqubitsprobanot
_{\Yqubitbothtimefinalstatecoefindexindexstd xx}
}

\newcommand{\Yqubitindexstd}
{i}
\newcommand{\ytwoqubitresultphaseinit}
{$\Ytwoqubitresultphaseinit$}
\newcommand{\Ytwoqubitresultphaseinit}
{\Delta _I}
\newcommand{\ytwoqubitresultphaseevol}
{$\Ytwoqubitresultphaseevol$}
\newcommand{\Ytwoqubitresultphaseevol}
{\Delta _E}
\newcommand{\ytwoqubitresultphaseevolindexd}
{$\Ytwoqubitresultphaseevolindexd$}
\newcommand{\Ytwoqubitresultphaseevolindexd}
{\Delta _{Ed}}
\newcommand{\ytwoqubitresultphaseevolindexdestim}
{$\Ytwoqubitresultphaseevolindexdestim$}
\newcommand{\Ytwoqubitresultphaseevolindexdestim}
{\widehat{\Delta} _{Ed}}
\newcommand{\ytwoqubitresultphaseevolsin}
{$\Ytwoqubitresultphaseevolsin$}
\newcommand{\Ytwoqubitresultphaseevolsin}
{v}

\newcommand{\Ytwoqubitresultphaseevolsinestim}
{\overline{\Ytwoqubitresultphaseevolsin}}

\newcommand{\ytwoqubitresultphaseevolsinestimtwo}
{$\Ytwoqubitresultphaseevolsinestimtwo$}
\newcommand{\Ytwoqubitresultphaseevolsinestimtwo}
{\hat{\Ytwoqubitresultphaseevolsin}}
\newcommand{\Yparamqubitbothstateplusmodulusnot}
{r}
\newcommand{\yparamqubitonestateplusmodulus}
{$\Yparamqubitonestateplusmodulus$}
\newcommand{\Yparamqubitonestateplusmodulus}
{{\Yparamqubitbothstateplusmodulusnot}_1}
\newcommand{\yparamqubittwostateplusmodulus}
{$\Yparamqubittwostateplusmodulus$}
\newcommand{\Yparamqubittwostateplusmodulus}
{{\Yparamqubitbothstateplusmodulusnot}_2}
\newcommand{\yparamqubitindexstdstateplusmodulus}
{$\Yparamqubitindexstdstateplusmodulus$}
\newcommand{\Yparamqubitindexstdstateplusmodulus}
{{\Yparamqubitbothstateplusmodulusnot}_{\Yqubitindexstd}}
\newcommand{\Yparamqubitbothstateminusmodulusnot}
{q}
\newcommand{\yparamqubitonestateminusmodulus}
{$\Yparamqubitonestateminusmodulus$}
\newcommand{\Yparamqubitonestateminusmodulus}
{{\Yparamqubitbothstateminusmodulusnot}_1}
\newcommand{\yparamqubittwostateminusmodulus}
{$\Yparamqubittwostateminusmodulus$}
\newcommand{\Yparamqubittwostateminusmodulus}
{{\Yparamqubitbothstateminusmodulusnot}_2}
\newcommand{\Yparamqubitindexstdstateminusmodulus}
{{\Yparamqubitbothstateminusmodulusnot}_{\Yqubitindexstd}}
\newcommand{\Yparamqubitbothstateplusphasenot}
{\theta}
\newcommand{\yparamqubitonestateplusphase}
{$\Yparamqubitonestateplusphase$}
\newcommand{\Yparamqubitonestateplusphase}
{\Yparamqubitbothstateplusphasenot_1}
\newcommand{\yparamqubittwostateplusphase}
{$\Yparamqubittwostateplusphase$}
\newcommand{\Yparamqubittwostateplusphase}
{\Yparamqubitbothstateplusphasenot_2}
\newcommand{\yparamqubitindexstdstateplusphase}
{$\Yparamqubitindexstdstateplusphase$}
\newcommand{\Yparamqubitindexstdstateplusphase}
{{\Yparamqubitbothstateplusphasenot}_{\Yqubitindexstd}}
\newcommand{\Yparamqubitbothstateminusphasenot}
{\phi}
\newcommand{\yparamqubitonestateminusphase}
{$\Yparamqubitonestateminusphase$}
\newcommand{\Yparamqubitonestateminusphase}
{\Yparamqubitbothstateminusphasenot_1}
\newcommand{\yparamqubittwostateminusphase}
{$\Yparamqubittwostateminusphase$}
\newcommand{\Yparamqubittwostateminusphase}
{\Yparamqubitbothstateminusphasenot_2}
\newcommand{\yparamqubitindexstdstateminusphase}
{$\Yparamqubitindexstdstateminusphase$}
\newcommand{\Yparamqubitindexstdstateminusphase}
{{\Yparamqubitbothstateminusphasenot}_{\Yqubitindexstd}}
\newcommand{\ymagfieldnot}
{$\Ymagfieldnot$}
\newcommand{\Ymagfieldnot}
{B}
\newcommand{\Ymagfieldvec}
{\overrightarrow{\Ymagfieldnot}}
\newcommand{\Yhamiltonfieldscale}
{G}
\newcommand{\yexchangetensorppalvaluexy}
{$\Yexchangetensorppalvaluexy$}
\newcommand{\Yexchangetensorppalvaluexy}
{J_{xy}}

\newcommand{\yexchangetensorppalvaluexyestim}
{$\Yexchangetensorppalvaluexyestim$}
\newcommand{\Yexchangetensorppalvaluexyestim}
{\widehat{J}_{xy}}
\newcommand{\yexchangetensorppalvaluexyestimargtwoqubitwritereadtimeintervalindexone}
{$\Yexchangetensorppalvaluexyestimargtwoqubitwritereadtimeintervalindexone$}
\newcommand{\Yexchangetensorppalvaluexyestimargtwoqubitwritereadtimeintervalindexone}
{\Yexchangetensorppalvaluexyestim
(
\Ytwoqubitwritereadtimeintervalindexone
)}
\newcommand{\yexchangetensorppalvaluexyshiftindetermint}
{$\Yexchangetensorppalvaluexyshiftindetermint$}
\newcommand{\Yexchangetensorppalvaluexyshiftindetermint}
{k_{xy}}
\newcommand{\yexchangetensorppalvaluexyshiftindetermintestim}
{$\Yexchangetensorppalvaluexyshiftindetermintestim$}
\newcommand{\Yexchangetensorppalvaluexyshiftindetermintestim}
{\widehat{k}_{xy}}
\newcommand{\yexchangetensorppalvaluexyshiftindetermintestimminusactual}
{$\Yexchangetensorppalvaluexyshiftindetermintestimminusactual$}
\newcommand{\Yexchangetensorppalvaluexyshiftindetermintestimminusactual}
{\Delta k _{xy}}
\newcommand{\yexchangetensorppalvaluez}
{$\Yexchangetensorppalvaluez$}
\newcommand{\Yexchangetensorppalvaluez}
{J_z}

\newcommand{\yexchangetensorppalvaluezestim}
{$\Yexchangetensorppalvaluezestim$}
\newcommand{\Yexchangetensorppalvaluezestim}
{\widehat{J}_z}
\newcommand{\yexchangetensorppalvaluezestimargtwoqubitwritereadtimeintervalindextwo}
{$\Yexchangetensorppalvaluezestimargtwoqubitwritereadtimeintervalindextwo$}
\newcommand{\Yexchangetensorppalvaluezestimargtwoqubitwritereadtimeintervalindextwo}
{\Yexchangetensorppalvaluezestim
(
\Ytwoqubitwritereadtimeintervalindextwo
)}
\newcommand{\yexchangetensorppalvaluezshiftindetermint}
{$\Yexchangetensorppalvaluezshiftindetermint$}
\newcommand{\Yexchangetensorppalvaluezshiftindetermint}
{k_{z}}
\newcommand{\yexchangetensorppalvaluezshiftindetermintestim}
{$\Yexchangetensorppalvaluezshiftindetermintestim$}
\newcommand{\Yexchangetensorppalvaluezshiftindetermintestim}
{\widehat{k}_{z}}
\newcommand{\yexchangetensorppalvaluezshiftindetermintestimminusactual}
{$\Yexchangetensorppalvaluezshiftindetermintestimminusactual$}
\newcommand{\Yexchangetensorppalvaluezshiftindetermintestimminusactual}
{\Delta k _{z}}
\newcommand{\ytwoqubitsprobadirxxplusplussumminusminusphasediff}
{$\Ytwoqubitsprobadirxxplusplussumminusminusphasediff$}
\newcommand{\Ytwoqubitsprobadirxxplusplussumminusminusphasediff}
{\Delta \Phi_{1,-1}}
\newcommand{\Ytwoqubitsprobadirxxplusplusdiffminusminusfactorreal}
{R_{14}}
\newcommand{\Ytwoqubitsprobadirxxplusplusdiffminusminusfactorimag}
{I_{14}}
\newcommand{\ytwoqubitsprobadirxxplusplusdiffminusminusphasediff}
{$\Ytwoqubitsprobadirxxplusplusdiffminusminusphasediff$}
\newcommand{\Ytwoqubitsprobadirxxplusplusdiffminusminusphasediff}
{\Delta \Phi_{1,0}}
\newcommand{\ytwoqubitsprobadirxxplusplusdiffminusminusphasediffindexd}
{$\Ytwoqubitsprobadirxxplusplusdiffminusminusphasediffindexd$}
\newcommand{\Ytwoqubitsprobadirxxplusplusdiffminusminusphasediffindexd}
{\Delta \Phi_{1,0d}}
\newcommand{\ytwoqubitsprobadirxxplusplusdiffminusminusphasediffindexdestim}
{$\Ytwoqubitsprobadirxxplusplusdiffminusminusphasediffindexdestim$}
\newcommand{\Ytwoqubitsprobadirxxplusplusdiffminusminusphasediffindexdestim}
{\widehat{\Delta \Phi}_{1,0d}}
\newcommand{\ytwoqubitsprobadirxxplusplusdiffminusminusphasediffcosgen}
{$\Ytwoqubitsprobadirxxplusplusdiffminusminusphasediffcosgen$}
\newcommand{\Ytwoqubitsprobadirxxplusplusdiffminusminusphasediffcosgen}
{w_1}
\newcommand{\ytwoqubitsprobadirxxplusplusdiffminusminusphasediffcosgenestim}
{$\Ytwoqubitsprobadirxxplusplusdiffminusminusphasediffcosgenestim$}
\newcommand{\Ytwoqubitsprobadirxxplusplusdiffminusminusphasediffcosgenestim}
{\widehat{w}_1}
\newcommand{\ytwoqubitsprobadirxxplusplusdiffminusminusphasediffsingen}
{$\Ytwoqubitsprobadirxxplusplusdiffminusminusphasediffsingen$}
\newcommand{\Ytwoqubitsprobadirxxplusplusdiffminusminusphasediffsingen}
{w_2}
\newcommand{\ytwoqubitsprobadirxxplusplusdiffminusminusphasediffsingenestim}
{$\Ytwoqubitsprobadirxxplusplusdiffminusminusphasediffsingenestim$}
\newcommand{\Ytwoqubitsprobadirxxplusplusdiffminusminusphasediffsingenestim}
{\widehat{w}_2}
\newcommand{\ysqrtminusone}
{$\Ysqrtminusone$}
\newcommand{\Ysqrtminusone}
{{\rm i}}
\newcommand{\ywritereadonestatenb}
{$\Ywritereadonestatenb$}
\newcommand{\Ywritereadonestatenb}
{K}

%% file: artiti106v1_eqdef.tex
\newcommand{\yeqdefhamiltonian}
{
\begin{eqnarray}
\nonumber
H
&
=
&
\Yhamiltonfieldscale s_{1z} \Ymagfieldnot
+ \Yhamiltonfieldscale s_{2z} \Ymagfieldnot
- 2 J_{xy} ( s_{1x} s_{2x} + s_{1y} s_{2y} )
\\
&
&
- 2 J_{z} s_{1z} s_{2z}
\label{eq-online-hamiltonian}
\end{eqnarray}
}
\newcommand{\yeqdeftwoqubitstateinitindexi}
{
\begin{equation}
\label{eq-twoqubit-state-init-index-i}
| \psi_i 
(
\Yqubitonetimeinit
)
\rangle
=
\alpha _i | + 
{\rangle}
+ \beta _i | - 
{\rangle}
\end{equation}
}
\newcommand{\yeqdefqubitpolarqubitindexstd}
{
\begin{equation}
\label{eq-def-qubit-polar-qubit-indexstd} 
\alpha_i
=
\Yparamqubitindexstdstateplusmodulus
e ^{ \Ysqrtminusone
\Yparamqubitindexstdstateplusphase }
\hspace{5mm} 
\beta _i
=
\Yparamqubitindexstdstateminusmodulus e ^{ \Ysqrtminusone
\Yparamqubitindexstdstateminusphase }
\hspace{10mm} 
\Yqubitindexstd \in \{ 1 , 2 \}
\end{equation}
}
\newcommand{\yeqdefparamqubitindexstdstateminusmodulusvsparamqubitindexstdstateplusmodulus}
{
\begin{equation}
\Yparamqubitindexstdstateminusmodulus
=
\sqrt
{
1
-
\Yparamqubitindexstdstateplusmodulus
^2
}
\hspace{10mm} 
\Yqubitindexstd \in \{ 1 , 2 \}
\label{eq-paramqubitindexstdstateminusmodulus-vs-paramqubitindexstdstateplusmodulus}
\end{equation}
}
\newcommand{\yeqdefqubitbothtimeinitstatetensorprod}
{
\begin{eqnarray}
\label{eq-qubitbothtimeinitstate-tensor-prod}
\Ymixsyststateinitial
&
=
&
\Yqubitonetimeinitstateindexone
\otimes
\Yqubitonetimeinitstateindextwo
\\
\nonumber
&
=
&
\alpha _1
\alpha _2
| ++ 
\rangle
+
\alpha _1
\beta _2
| +- 
\rangle
\\
\label{eq-etat-deuxspin-plusplus-initial-decompos}
&
&
+
\beta _1
\alpha _2
| -+ 
\rangle
+
\beta _1
\beta _2
| -- 
\rangle
\end{eqnarray}
}
\newcommand{\yeqdefstatetfinalvsopmixstatetinitcomponents}
{
\begin{equation}
\label{eq-statetfinalvsopmixstatetinit-components}
\Yveccompsyststatetfinal
=
\Yopmix
\Yveccompsyststatetinit
\end{equation}
}
\newcommand{\yeqdefveccompsyststatetinit}
{
\begin{equation}
\Yveccompsyststatetinit
=
[
\alpha _1
\alpha _2
,
\alpha _1
\beta _2
,
\beta _1
\alpha _2
,
\beta _1
\beta _2
]
^T
\end{equation}
}
\newcommand{\yeqdefopmixmatrixdecompose}
{
\begin{equation}
\label{eq-opmixmatrixdecompose}
\Yopmix 
= \Yopmixbases \Yopmixdiag \Yopmixbases^{-1}
= \Yopmixbases \Yopmixdiag \Yopmixbases
\end{equation}
}
\newcommand{\yeqdefopmixbasesdef}
{
\begin{equation}
\label{eq-opmixbasesdef}
\Yopmixbases 
=
\Yopmixbases ^{-1}
=
\left[
\begin{tabular}{llll}
1
&
0
&
0
&
0
\\
0
&
$\frac{1}{\sqrt{2}}$
&
$\frac{1}{\sqrt{2}}$
&
0
\\
0
&
$\frac{1}{\sqrt{2}}$
&
$- \frac{1}{\sqrt{2}}$
&
0
\\
0
&
0
&
0
&
1
\end{tabular}
\right]
\end{equation}
}
\newcommand{\yeqdefopmixdiagdef}
{
{ 
\begin{eqnarray}
\left[
\begin{tabular}{llll}
$
e^{
-
\Ysqrtminusone
\omega _{1 , 1}
(t - t_0)
}
$
&
0
&
0
&
0
\\
0
&
$
e^{
- 
\Ysqrtminusone
\omega _{1 , 0}
(t - t_0)
}
$
&
0
&
0
\\
0
&
0&
$
e^{
- 
\Ysqrtminusone
\omega _{0 , 0}
(t - t_0)
}
$
&
0
\\
0
&
0
&
0
&
$
e^{
- 
\Ysqrtminusone
\omega _{1 , -1}
(t - t_0)
}
$
\end{tabular}
\right]
.
\nonumber
\\
\label{eq-opmixdiagdef}
\end{eqnarray}
}
}
\newcommand{\yeqdefomegaallexpress}
{
\begin{eqnarray}
\hspace{-3mm}
\omega _{1, 1}
=
\frac{1}{\hbar}
\left[
\Yhamiltonfieldscale \Ymagfieldnot - 
\frac{
\Yexchangetensorppalvaluez
}
{2}
\right]
,
&
\hspace{3mm}
&
\label{eq-omega-one-zero-express}
\omega _{1, 0}
=
\frac{1}{\hbar}
\left[
- 
\Yexchangetensorppalvaluexy
+ 
\frac{
\Yexchangetensorppalvaluez
}
{2}
\right]
,
\\
\label{eq-omega-zero-zero-express}
\hspace{-33mm}
\omega _{0, 0}
=
\frac{1}{\hbar}
\left[
\Yexchangetensorppalvaluexy
+ 
\frac{ 
\Yexchangetensorppalvaluez
}
{2}
\right]
,
&
\hspace{3mm}
&
\omega _{1, - 1}
=
\frac{1}{\hbar}
\left[
- \Yhamiltonfieldscale
\Ymagfieldnot - 
\frac{ 
\Yexchangetensorppalvaluez
}
{2}
\right]
.
\end{eqnarray}
}
\newcommand{\yeqdefstatecoefvstwoqubitsprobathreepolar}
{
\begin{eqnarray}
\label{eq-statecoef-vs-twoqubitsprobaplusplus-polar}
\Ytwoqubitsprobaplusplusdirzz
&
=
& 
\Yparamqubitonestateplusmodulus ^2 \Yparamqubittwostateplusmodulus
^2 
\\
\nonumber
\Ytwoqubitsprobaplusminusdirzz
&
=
&
\Yparamqubitonestateplusmodulus ^2 ( 1
-
\Yparamqubittwostateplusmodulus ^2 ) ( 1 -
\Ytwoqubitresultphaseevolsin ^2 ) + ( 1
-
\Yparamqubitonestateplusmodulus ^2 )
\Yparamqubittwostateplusmodulus ^2 \Ytwoqubitresultphaseevolsin ^2
\\
&
&
{
-
2 \Yparamqubitonestateplusmodulus \Yparamqubittwostateplusmodulus
\sqrt{ 1
-
\Yparamqubitonestateplusmodulus ^2 } \sqrt{ 1
-
\Yparamqubittwostateplusmodulus ^2 } \sqrt{ 1 -
\Ytwoqubitresultphaseevolsin ^2 } \Ytwoqubitresultphaseevolsin
\sin \Ytwoqubitresultphaseinit
}
\label{eq-statecoef-vs-twoqubitsprobaplusminus-polar-versionthree}
\\
\label{eq-statecoef-vs-twoqubitsprobaminusminus-polar-vs-paramqubitindexstdstateplusmodulus}
\Ytwoqubitsprobaminusminusdirzz
&
=
&
( 1
-
\Yparamqubitonestateplusmodulus ^2 ) ( 1
-
\Yparamqubittwostateplusmodulus ^2 )
\end{eqnarray}
}
\newcommand{\yeqdefeqdeftwoqubitresultphaseinittwoqubitresultphaseevol}
{
\begin{eqnarray}
\label{eq-def-twoqubitresultphaseinit-twoqubitresultphaseevol}
\Ytwoqubitresultphaseinit
&
=
&
( \Yparamqubittwostateminusphase
-
\Yparamqubittwostateplusphase
)
-
(
\Yparamqubitonestateminusphase
-
\Yparamqubitonestateplusphase )
\\
\label{eq-def-Ytwoqubitresultphaseevol-versiontwo}
\Ytwoqubitresultphaseevol
&
=
&
- 
\frac{ 
\Yexchangetensorppalvaluexy
( t - t_0 )
} { \hbar } 
\\
\Ytwoqubitresultphaseevolsin
&
=
&
\mbox{sgn} ( \cos \Ytwoqubitresultphaseevol )
\sin \Ytwoqubitresultphaseevol
.
\label{eq-def-Ytwoqubitresultphaseevol-versionthree}
\end{eqnarray}
}

\newcommand{\yeqdeftwoqubitsprobaplusplusdirxxminustwoqubitsprobaminusminusdirxxdefandexplicit}
{
\begin{equation}
\Ytwoqubitsprobaplusplusdirxx
-
\Ytwoqubitsprobaminusminusdirxx
=
\Ytwoqubitsprobadirxxplusplusdiffminusminusfactorreal
\Ytwoqubitsprobadirxxplusplusdiffminusminusphasediffcosgen
-
\Ytwoqubitsprobadirxxplusplusdiffminusminusfactorimag
\Ytwoqubitsprobadirxxplusplusdiffminusminusphasediffsingen
\label{eq-twoqubitsprobaplusplusdirxx-minus-twoqubitsprobaminusminusdirxx-def-and-explicit}
\end{equation}
}
\newcommand{\yeqdeftwoqubitsprobadirxxplusplusdiffminusminusphasediffvsexchangetensor}
{
\begin{eqnarray}
\Ytwoqubitsprobadirxxplusplusdiffminusminusfactorreal
&
=
&
\Yparamqubitonestateplusmodulus
^2
\Yparamqubittwostateplusmodulus
\sqrt{
1
-
\Yparamqubittwostateplusmodulus
^2
}
\cos
( \Yparamqubittwostateminusphase
-
\Yparamqubittwostateplusphase
)
\nonumber
\\
&
&
+
\Yparamqubittwostateplusmodulus
^2
\Yparamqubitonestateplusmodulus
\sqrt{
1
-
\Yparamqubitonestateplusmodulus
^2
}
\cos
(
\Yparamqubitonestateminusphase
-
\Yparamqubitonestateplusphase )
\nonumber
\\
&
&
+
(
1
-
\Yparamqubitonestateplusmodulus
^2
)
\Yparamqubittwostateplusmodulus
\sqrt{
1
-
\Yparamqubittwostateplusmodulus
^2
}
\cos
( \Yparamqubittwostateminusphase
-
\Yparamqubittwostateplusphase
-
\Ytwoqubitsprobadirxxplusplussumminusminusphasediff
)
\nonumber
\\
&
&
+
(
1
-
\Yparamqubittwostateplusmodulus
^2
)
\Yparamqubitonestateplusmodulus
\sqrt{
1
-
\Yparamqubitonestateplusmodulus
^2
}
\cos
(
\Yparamqubitonestateminusphase
-
\Yparamqubitonestateplusphase
-
\Ytwoqubitsprobadirxxplusplussumminusminusphasediff
)
\nonumber
\\
\label{eq-twoqubitsprobadirxxplusplusdiffminusminusfactorreal-explicit}
\end{eqnarray}
\begin{eqnarray}
\Ytwoqubitsprobadirxxplusplusdiffminusminusfactorimag
&
=
&
-
\Yparamqubitonestateplusmodulus
^2
\Yparamqubittwostateplusmodulus
\sqrt{
1
-
\Yparamqubittwostateplusmodulus
^2
}
\sin
( \Yparamqubittwostateminusphase
-
\Yparamqubittwostateplusphase
)
\nonumber
\\
&
&
-
\Yparamqubittwostateplusmodulus
^2
\Yparamqubitonestateplusmodulus
\sqrt{
1
-
\Yparamqubitonestateplusmodulus
^2
}
\sin
(
\Yparamqubitonestateminusphase
-
\Yparamqubitonestateplusphase )
\nonumber
\\
&
&
+
(
1
-
\Yparamqubitonestateplusmodulus
^2
)
\Yparamqubittwostateplusmodulus
\sqrt{
1
-
\Yparamqubittwostateplusmodulus
^2
}
\sin
( \Yparamqubittwostateminusphase
-
\Yparamqubittwostateplusphase
-
\Ytwoqubitsprobadirxxplusplussumminusminusphasediff
)
\nonumber
\\
&
&
+
(
1
-
\Yparamqubittwostateplusmodulus
^2
)
\Yparamqubitonestateplusmodulus
\sqrt{
1
-
\Yparamqubitonestateplusmodulus
^2
}
\sin
(
\Yparamqubitonestateminusphase
-
\Yparamqubitonestateplusphase
-
\Ytwoqubitsprobadirxxplusplussumminusminusphasediff
)
\nonumber
\\
\label{eq-twoqubitsprobadirxxplusplusdiffminusminusfactorimag-explicit}
\\
\Ytwoqubitsprobadirxxplusplusdiffminusminusphasediffcosgen
&
=
&
\cos
\Ytwoqubitsprobadirxxplusplusdiffminusminusphasediff
\label{eq-def-twoqubitsprobadirxxplusplusdiffminusminusphasediffcosgen}
\\
\Ytwoqubitsprobadirxxplusplusdiffminusminusphasediffsingen
&
=
&
\sin
\Ytwoqubitsprobadirxxplusplusdiffminusminusphasediff
\label{eq-def-twoqubitsprobadirxxplusplusdiffminusminusphasediffsingen}
\\
\Ytwoqubitsprobadirxxplusplussumminusminusphasediff
&
=
&
-
\frac{
2
\Yhamiltonfieldscale
\Ymagfieldnot
(t - t_0)
}
{
\hbar
}
\label{eq-combine-omegaoneone-omegaoneminusone}
\\
\Ytwoqubitsprobadirxxplusplusdiffminusminusphasediff
&
=
&
\frac{
(t - t_0)
}
{
\hbar
}
\left(
-
J_{xy}
+
J_{z}
-
\Yhamiltonfieldscale
\Ymagfieldnot
\right)
.
\label{eq-twoqubitsprobadirxxplusplusdiffminusminusphasediff-vs-exchange-tensor}
\end{eqnarray}
}
\newcommand{\yeqdeftwoqubitsprobaplusplusdirxxplustwoqubitsprobaminusminusdirxx}
{
\begin{eqnarray}
\Ytwoqubitsprobaplusplusdirxx
+
\Ytwoqubitsprobaminusminusdirxx
&
=
&
\frac{1}{2}
+
\Yparamqubitonestateplusmodulus
\Yparamqubittwostateplusmodulus
\sqrt{
1
-
\Yparamqubitonestateplusmodulus
^2
}
\sqrt{
1
-
\Yparamqubittwostateplusmodulus
^2
}
\left[
\cos
\Ytwoqubitresultphaseinit
\right.
\nonumber
\\
&
&
\hspace{3mm}
\left.
+
\cos
\left(
(
\Yparamqubitonestateminusphase
-
\Yparamqubitonestateplusphase )
+
( \Yparamqubittwostateminusphase
-
\Yparamqubittwostateplusphase
)
-
\Ytwoqubitsprobadirxxplusplussumminusminusphasediff
\right)
\right]
.
\nonumber
\\
\label{eq-twoqubitsprobaplusplusdirxx-plus-twoqubitsprobaminusminusdirxx-explicit}
\end{eqnarray}
}
\newcommand{\yeqdefstatecoefvstwoqubitsprobaplusminuspolarversionthreeexpect}
{
\begin{eqnarray}
E
\{
\Ytwoqubitsprobaplusminusdirzz
\}
&
=
&
E
\{
\Yparamqubitonestateplusmodulus ^2 
\}
( 1
-
E
\{
\Yparamqubittwostateplusmodulus ^2 
\}
) ( 1 -
\Ytwoqubitresultphaseevolsin ^2 ) 
\nonumber
\\
&
&
+ ( 1
-
E
\{
\Yparamqubitonestateplusmodulus ^2 
\}
)
E
\{
\Yparamqubittwostateplusmodulus ^2 
\}
\Ytwoqubitresultphaseevolsin ^2
\nonumber
\\
&
&
-
2 
E
\{
\Yparamqubitonestateplusmodulus 
\sqrt{ 1
-
\Yparamqubitonestateplusmodulus ^2 } 
\}
E
\{
\Yparamqubittwostateplusmodulus
\sqrt{ 1
-
\Yparamqubittwostateplusmodulus ^2 } 
\}
\sqrt{ 1 -
\Ytwoqubitresultphaseevolsin ^2 } \Ytwoqubitresultphaseevolsin
\nonumber
\\
&
&
\hspace{2mm}
\times
E
\{
\sin \Ytwoqubitresultphaseinit
\}
.
\label{eq-statecoef-vs-twoqubitsprobaplusminus-polar-versionthree-expect}
\end{eqnarray}
}
\newcommand{\yeqdefparamqubitonestateplusmodulusltonehalfltparamqubittwostateplusmodulus}
{
\begin{equation}
\label{eq-qubit-modulus-allowed-range}
0 
<
\Yparamqubitonestateplusmodulus < \frac{1}{2} <
\Yparamqubittwostateplusmodulus 
<
1 
\end{equation}
}
\newcommand{\yeqdeftwoqubitresultphaseevolsinvsfirstordermomentsqr}
{
\begin{equation}
\Ytwoqubitresultphaseevolsin
^2
=
\frac{
E \{
\Ytwoqubitsprobaplusminusdirzz
\}
-
E \{ \Yparamqubitonestateplusmodulus ^2 \} ( 1
-
E \{ \Yparamqubittwostateplusmodulus ^2 \} )
}
{
{%
E \{ \Yparamqubittwostateplusmodulus ^2 \}
-
E \{ \Yparamqubitonestateplusmodulus ^2 \}
}
}
.
\label{eq-twoqubitresultphaseevolsin-vs-first-order-moment-sqr}
\end{equation}
}
\newcommand{\yeqdefqubitindexstdstateplusmodulussolbothfrommlsp}
{
\begin{eqnarray}
\Yparamqubitindexstdstateplusmodulus
&
=
&
\left[
\frac{1}{2}
\left[ (1 + 
\Ytwoqubitsprobaplusplusdirzz
- 
\Ytwoqubitsprobaminusminusdirzz
)
\right.
\right.
\nonumber
\\
&
&
\left.
\left.
+
\epsilon
_{\Yqubitindexstd}
\sqrt{ (1 + 
\Ytwoqubitsprobaplusplusdirzz
- 
\Ytwoqubitsprobaminusminusdirzz
)
^2 - 4 
\Ytwoqubitsprobaplusplusdirzz
}
\right]
\right]
^
{1/2}
\hspace{5mm}
\Yqubitoneindexstd
\in
\{
1,
2
\}
\nonumber
\\
\label{eq-qubitindexstdstateplusmodulus-sol-both-from-mlsp}
\end{eqnarray}
}
\newcommand{\yeqdefstatecoefvstwoqubitsprobathreepolarmean}
{
\begin{eqnarray}
\label{eq-statecoef-vs-twoqubitsprobaplusplus-polar-mean}
E
\{
\Ytwoqubitsprobaplusplusdirzz
\}
&
=
& 
E
\{
\Yparamqubitonestateplusmodulus ^2 
\}
E
\{
\Yparamqubittwostateplusmodulus
^2 
\}
\\
\label{eq-statecoef-vs-twoqubitsprobaminusminus-polar-vs-paramqubitindexstdstateplusmodulus-mean}
E
\{
\Ytwoqubitsprobaminusminusdirzz
\}
&
=
&
( 1
-
E
\{
\Yparamqubitonestateplusmodulus ^2 
\}
) ( 1
-
E
\{
\Yparamqubittwostateplusmodulus ^2 
\}
)
.
\end{eqnarray}
}
\newcommand{\yeqdefparamqubitindexstdstateplusmodulussolbothdirzz}
{
\begin{eqnarray}
E
\{
\Yparamqubitindexstdstateplusmodulus
^2
\}
&
=
&
\frac{1}{2}
\left[ (1 + 
E
\{
\Ytwoqubitsprobaplusplusdirzz
\}
- 
E
\{
\Ytwoqubitsprobaminusminusdirzz
\}
)
\right.
\nonumber
\\
&
&
\left.
+
\epsilon
_{\Yqubitindexstd}
\sqrt{ (1 + 
E
\{
\Ytwoqubitsprobaplusplusdirzz
\}
- 
E
\{
\Ytwoqubitsprobaminusminusdirzz
\}
)
^2 - 4 
E
\{
\Ytwoqubitsprobaplusplusdirzz
\}
}
\right]
\nonumber
\\
&
&
\hspace{10mm}
\Yqubitoneindexstd
\in
\{
1,
2
\}
\label{eq-paramqubitindexstdstateplusmodulus-sol-both}
\end{eqnarray}
}

\newcommand{\yeqdeftwoqubitresultphaseevolsinvsitssqr}
{
\begin{equation}
\Ytwoqubitresultphaseevolsin
=
\frac{
E
\{
\Yparamqubitonestateplusmodulus ^2 
\}
( 1
-
E
\{
\Yparamqubittwostateplusmodulus ^2 
\}
)
+
(
E
\{
\Yparamqubittwostateplusmodulus ^2 
\}
-
E
\{
\Yparamqubitonestateplusmodulus ^2 
\}
)
\Ytwoqubitresultphaseevolsin ^2
-
E
\{
\Ytwoqubitsprobaplusminusdirzz
\}
}
{
2 
E
\{
\Yparamqubitonestateplusmodulus 
\sqrt{ 1
-
\Yparamqubitonestateplusmodulus ^2 } 
\}
E
\{
\Yparamqubittwostateplusmodulus
\sqrt{ 1
-
\Yparamqubittwostateplusmodulus ^2 } 
\}
\sqrt{ 1 -
\Ytwoqubitresultphaseevolsin ^2 } 
E
\{
\sin \Ytwoqubitresultphaseinit
\}
}
.
\label{eq-twoqubitresultphaseevolsin-vs-its-sqr}
\end{equation}
}
\newcommand{\yeqdeftwoqubitresultphaseevolsinvsitssqrsgn}
{
\begin{eqnarray}
\mathrm{sgn}
(
\Ytwoqubitresultphaseevolsin
)
&
=
&
\mathrm{sgn}
\left(
E
\{
\Yparamqubitonestateplusmodulus ^2 
\}
( 1
-
E
\{
\Yparamqubittwostateplusmodulus ^2 
\}
)
+
(
E
\{
\Yparamqubittwostateplusmodulus ^2 
\}
-
E
\{
\Yparamqubitonestateplusmodulus ^2 
\}
)
\Ytwoqubitresultphaseevolsin ^2
\right.
\nonumber
\\
&
&
\left.
\hspace{7mm}
-
E
\{
\Ytwoqubitsprobaplusminusdirzz
\}
\right)
\mathrm{sgn}
(
E
\{
\sin \Ytwoqubitresultphaseinit
\}
)
.
\label{eq-twoqubitresultphaseevolsin-vs-its-sqr-sgn}
\end{eqnarray}
}
\newcommand{\yeqdefparamqubitindexstdstateplusandminusphaseconstraints}
{
\begin{eqnarray}
E
\{
\sin
(
\Yparamqubitindexstdstateminusphase
-
\Yparamqubitindexstdstateplusphase
)
\}
&
=
&
0
\hspace{10mm}
\Yqubitindexstd
\in
\{
1
,
2
\}
\\
E
\{
\cos
(
\Yparamqubitindexstdstateminusphase
-
\Yparamqubitindexstdstateplusphase
)
\}
&
>
&
0
\hspace{10mm}
\Yqubitindexstd
\in
\{
1
,
2
\}
,
\label{eq-cos-paramqubitindexstdstateminusminusplusphase-positive}
\end{eqnarray}
}
\newcommand{\yeqdeftwoqubitsprobaplusplusdirzzmeanandothers}
{
\begin{eqnarray}
E
\{
\Ytwoqubitsprobaplusplusdirzz
\}
&
=
&
(
E
\{
\Yparamqubitindexstdstateplusmodulus
^2
\}
)
^2
\label{eq-twoqubitsprobaplusplusdirzz-mean}
\\
E
\{
\Ytwoqubitsprobaplusplusdirxx
\}
-
E
\{
\Ytwoqubitsprobaminusminusdirxx
\}
&
=
&
E
\{
\Ytwoqubitsprobadirxxplusplusdiffminusminusfactorreal
\}
\Ytwoqubitsprobadirxxplusplusdiffminusminusphasediffcosgen
-
E
\{
\Ytwoqubitsprobadirxxplusplusdiffminusminusfactorimag
\}
\Ytwoqubitsprobadirxxplusplusdiffminusminusphasediffsingen
\label{eq-twoqubitsprobaplusplusdirxx-minus-twoqubitsprobaminusminusdirxx-def-and-explicit-mean}
\end{eqnarray}
\begin{eqnarray}
E
\{
\Ytwoqubitsprobadirxxplusplusdiffminusminusfactorreal
\}
&
=
&
E
\{
\Yparamqubitindexstdstateplusmodulus
\sqrt{
1
-
\Yparamqubitindexstdstateplusmodulus
^2
}
\}
E
\{
\cos
(
\Yparamqubitindexstdstateminusphase
-
\Yparamqubitindexstdstateplusphase
)
\}
\nonumber
\\
&
&
\hspace{3mm}
\times
2
\left[
E
\{
\Yparamqubitindexstdstateplusmodulus
^2
\}
(
1
-
\cos
\Ytwoqubitsprobadirxxplusplussumminusminusphasediff
)
+
\cos
\Ytwoqubitsprobadirxxplusplussumminusminusphasediff
\right]
\nonumber
\\
&
&
\label{eq-twoqubitsprobadirxxplusplusdiffminusminusfactorreal-explicit-mean}
\\
E
\{
\Ytwoqubitsprobadirxxplusplusdiffminusminusfactorimag
\}
&
=
&
-
E
\{
\Yparamqubitindexstdstateplusmodulus
\sqrt{
1
-
\Yparamqubitindexstdstateplusmodulus
^2
}
\}
E
\{
\cos
(
\Yparamqubitindexstdstateminusphase
-
\Yparamqubitindexstdstateplusphase
)
\}
\nonumber
\\
&
&
\hspace{3mm}
\times
2
(
1
-
E
\{
\Yparamqubitindexstdstateplusmodulus
^2
\}
)
\sin
\Ytwoqubitsprobadirxxplusplussumminusminusphasediff
\label{eq-twoqubitsprobadirxxplusplusdiffminusminusfactorimag-explicit-mean}
\end{eqnarray}
\begin{eqnarray}
E
\{
\Ytwoqubitsprobaplusplusdirxx
\}
+
E
\{
\Ytwoqubitsprobaminusminusdirxx
\}
&
=
&
\left[
E
\{
\Yparamqubitindexstdstateplusmodulus
\sqrt{
1
-
\Yparamqubitindexstdstateplusmodulus
^2
}
\}
E
\{
\cos
(
\Yparamqubitindexstdstateminusphase
-
\Yparamqubitindexstdstateplusphase
)
\}
\right]
^2
\nonumber
\\
&
&
\times
(
1
+
\cos
\Ytwoqubitsprobadirxxplusplussumminusminusphasediff
)
+
\frac{1}{2}
.
\label{eq-twoqubitsprobaplusplusdirxx-plus-twoqubitsprobaminusminusdirxx-mean}
\end{eqnarray}
}
\newcommand{\yeqdeftwoqubitsprobaplusplusdirzzmeansqrt}
{
\begin{equation}
E
\{
\Yparamqubitindexstdstateplusmodulus
^2
\}
=
\sqrt{
E
\{
\Ytwoqubitsprobaplusplusdirzz
\}
}
.
\label{eq-twoqubitsprobaplusplusdirzz-mean-sqrt}
\end{equation}
}
\newcommand{\yeqdeftwoqubitsprobaplusplusdirxxplustwoqubitsprobaminusminusdirxxmeanresultingstat}
{
\begin{equation}
E
\{
\Yparamqubitindexstdstateplusmodulus
\sqrt{
1
-
\Yparamqubitindexstdstateplusmodulus
^2
}
\}
E
\{
\cos
(
\Yparamqubitindexstdstateminusphase
-
\Yparamqubitindexstdstateplusphase
)
\}
=
\left[
\frac
{
E
\{
\Ytwoqubitsprobaplusplusdirxx
\}
+
E
\{
\Ytwoqubitsprobaminusminusdirxx
\}
-
\frac{1}{2}
}
{
1
+
\cos
\Ytwoqubitsprobadirxxplusplussumminusminusphasediff
}
\right]
^{\frac{1}{2}}
.
\label{eq-twoqubitsprobaplusplusdirxx-plus-twoqubitsprobaminusminusdirxx-mean-resulting-stat}
\end{equation}
}
\newcommand{\yeqdefexchangetensorppalvaluexyvstwoqubitresultphaseevolindexd}
{
\begin{equation}
\frac{ 
\Yexchangetensorppalvaluexy
\Ytwoqubitwritereadtimeintervalindexone
} 
{ \hbar }
=
-
\Ytwoqubitresultphaseevolindexd
+
\Yexchangetensorppalvaluexyshiftindetermint
\pi
\label{eq-exchangetensorppalvaluexy-vs-twoqubitresultphaseevolindexd}
\end{equation}
with
\begin{equation}
\Ytwoqubitresultphaseevolindexd
=
\mathrm{arcsin}
( \Ytwoqubitresultphaseevolsin )
\label{eq-twoqubitresultphaseevolindexd}
\end{equation}
}
\newcommand{\yeqdefexchangetensorppalvaluezvstwoqubitsprobadirxxplusplusdiffminusminusphasediffindexd}
{
\begin{equation}
\frac{ 
\Yexchangetensorppalvaluez
\Ytwoqubitwritereadtimeintervalindexone
} 
{ \hbar }
=
\Ytwoqubitsprobadirxxplusplusdiffminusminusphasediffindexd
+
2
\Yexchangetensorppalvaluezshiftindetermint
\pi
+
\frac{ 
\Yexchangetensorppalvaluexy
\Ytwoqubitwritereadtimeintervalindexone
} 
{ \hbar }
+
\frac{ 
\Yhamiltonfieldscale
\Ymagfieldnot
\Ytwoqubitwritereadtimeintervalindexone
} 
{ \hbar }
\label{eq-exchangetensorppalvaluez-vs-twoqubitsprobadirxxplusplusdiffminusminusphasediffindexd}
\end{equation}
with
\begin{equation}
\Ytwoqubitsprobadirxxplusplusdiffminusminusphasediffindexd
=
\mathrm{sgn}
(
\Ytwoqubitsprobadirxxplusplusdiffminusminusphasediffsingen
)
\
\mathrm{arccos}
(
\Ytwoqubitsprobadirxxplusplusdiffminusminusphasediffcosgen
)
\label{eq-def-twoqubitsprobadirxxplusplusdiffminusminusphasediffindexd}
\end{equation}
}
\newcommand{\yeqdefexchangetensorppalvaluexyvstwoqubitresultphaseevolindexdestim}
{
\begin{equation}
\frac{ 
\Yexchangetensorppalvaluexyestim
\Ytwoqubitwritereadtimeintervalindexone
} 
{ \hbar }
=
-
\Ytwoqubitresultphaseevolindexdestim
+
\Yexchangetensorppalvaluexyshiftindetermintestim
\pi
.
\label{eq-exchangetensorppalvaluexy-vs-twoqubitresultphaseevolindexd-estim}
\end{equation}
}
\newcommand{\yeqdefexchangetensorppalvaluezvstwoqubitsprobadirxxplusplusdiffminusminusphasediffindexdestim}
{
\begin{equation}
\frac{ 
\Yexchangetensorppalvaluezestim
\Ytwoqubitwritereadtimeintervalindexone
} 
{ \hbar }
=
\Ytwoqubitsprobadirxxplusplusdiffminusminusphasediffindexdestim
+
2
\Yexchangetensorppalvaluezshiftindetermintestim
\pi
+
\frac{ 
\Yexchangetensorppalvaluexyestim
\Ytwoqubitwritereadtimeintervalindexone
} 
{ \hbar }
+
\frac{ 
\Yhamiltonfieldscale
\Ymagfieldnot
\Ytwoqubitwritereadtimeintervalindexone
} 
{ \hbar }
\label{eq-exchangetensorppalvaluez-vs-twoqubitsprobadirxxplusplusdiffminusminusphasediffindexd-estim}
\end{equation}
}
\newcommand{\yeqdefexchangetensorppalvaluebothestimvsactual}
{
\begin{eqnarray}
\frac{ 
\Yexchangetensorppalvaluexyestim
\Ytwoqubitwritereadtimeintervalindexone
} 
{ \hbar }
&
=
&
\frac{ 
\Yexchangetensorppalvaluexy
\Ytwoqubitwritereadtimeintervalindexone
} 
{ \hbar }
+
\Yexchangetensorppalvaluexyshiftindetermintestimminusactual
\pi
\label{eq-exchangetensorppalvaluexy-estim-vs-actual}
\\
\frac{ 
\Yexchangetensorppalvaluezestim
\Ytwoqubitwritereadtimeintervalindexone
} 
{ \hbar }
&
=
&
\frac{ 
\Yexchangetensorppalvaluez
\Ytwoqubitwritereadtimeintervalindexone
} 
{ \hbar }
+
2
\Yexchangetensorppalvaluezshiftindetermintestimminusactual
\pi
+
\Yexchangetensorppalvaluexyshiftindetermintestimminusactual
\pi
\label{eq-exchangetensorppalvaluez-estim-vs-actual}
\end{eqnarray}
with
\begin{eqnarray}
\Yexchangetensorppalvaluexyshiftindetermintestimminusactual
&
=
&
\Yexchangetensorppalvaluexyshiftindetermintestim
-
\Yexchangetensorppalvaluexyshiftindetermint
\\
\Yexchangetensorppalvaluezshiftindetermintestimminusactual
&
=
&
\Yexchangetensorppalvaluezshiftindetermintestim
-
\Yexchangetensorppalvaluezshiftindetermint
.
\end{eqnarray}
}
\newcommand{\yeqdefexchangetensorppalvaluebothestimvsactualargtwoqubitwritereadtimeintervalindexone}
{
\begin{equation}
\frac{ 
\Yexchangetensorppalvaluexyestimargtwoqubitwritereadtimeintervalindexone
\Ytwoqubitwritereadtimeintervalindexone
} 
{ \hbar }
=
\frac{ 
\Yexchangetensorppalvaluexy
\Ytwoqubitwritereadtimeintervalindexone
} 
{ \hbar }
+
\Yexchangetensorppalvaluexyshiftindetermintestimminusactual
\pi
.
\label{eq-exchangetensorppalvaluexy-estim-vs-actual-argtwoqubitwritereadtimeintervalindexone}
\end{equation}
}
\newcommand{\yeqdefopmixdiagestimvsopmixdiag}
{
\begin{equation}
\Yopmixdiagestim
=
e^{
\Ysqrtminusone
(
\Yexchangetensorppalvaluezshiftindetermintestimminusactual
\pi
-
\Yexchangetensorppalvaluexyshiftindetermintestimminusactual
\frac{\pi}{2}
)
}
\Yopmixdiag
.
\label{yeq-opmixdiagestim-vs-opmixdiag}
\end{equation}
}
\newcommand{\yeqdefexchangetensorppalvaluezestimvsactualtwoqubitwritereadtimeintervalindextwo}
{
\begin{equation}
\frac{ 
\ymodifartitionehundredsixvonestepone{\Yexchangetensorppalvaluezestimargtwoqubitwritereadtimeintervalindextwo}
\Ytwoqubitwritereadtimeintervalindextwo
} 
{ \hbar }
=
\frac{ 
\Yexchangetensorppalvaluez
\Ytwoqubitwritereadtimeintervalindextwo
} 
{ \hbar }
+
2
\Yexchangetensorppalvaluezshiftindetermintestimminusactual
\pi
+
2
\Yexchangetensorppalvaluexyshiftindetermintestimminusactual
\pi
.
\label{eq-exchangetensorppalvaluez-estim-vs-actual-twoqubitwritereadtimeintervalindextwo}
\end{equation}
}
\newcommand{\yeqdefexchangetensorppalvaluezestimvsactualtwoqubitwritereadtimeintervalindexthree}
{
\begin{eqnarray}
\frac{ 
\ymodifartitionehundredsixvonestepone{\Yexchangetensorppalvaluexyestimargtwoqubitwritereadtimeintervalindexone}
\Ytwoqubitwritereadtimeintervalindexthree
} 
{ \hbar }
&
=
&
\frac{ 
\Yexchangetensorppalvaluexy
\Ytwoqubitwritereadtimeintervalindexthree
} 
{ \hbar }
+
4
\Yexchangetensorppalvaluexyshiftindetermintestimminusactual
\pi
\label{eq-exchangetensorppalvaluexy-estim-vs-actual-twoqubitwritereadtimeintervalindexthree}
\\
\frac{ 
\ymodifartitionehundredsixvonestepone{\Yexchangetensorppalvaluezestimargtwoqubitwritereadtimeintervalindextwo}
\Ytwoqubitwritereadtimeintervalindexthree
} 
{ \hbar }
&
=
&
\frac{ 
\Yexchangetensorppalvaluez
\Ytwoqubitwritereadtimeintervalindexthree
} 
{ \hbar }
+
4
\Yexchangetensorppalvaluezshiftindetermintestimminusactual
\pi
+
4
\Yexchangetensorppalvaluexyshiftindetermintestimminusactual
\pi
.
\label{eq-exchangetensorppalvaluez-estim-vs-actual-twoqubitwritereadtimeintervalindexthree}
\end{eqnarray}
}

%% file: artiti106v1_core_for_upload.tex
\section{Introduction}\label{sec-intro}
System identification and system inversion are two closely related
problems. First considering classical, i.e. non-quantum, signals and
systems, the basic 
version of system identification 
concerns single-input single-output (SISO)
systems.
It
consists of
estimating the unknown parameter values of such a 
system
(i.e. of the transform that it performs)
belonging to a known class, by using known values of
its input 
(source signal
\ysssrcsigdisctimecentvec )
and output
(signal 
\yssmixsigdisctimecentvec ).
This version
\cite{book-ljung}
is
stated to be ``non-blind'' by the signal and image
processing community
\cite{a593}
or ``supervised'' by the machine learning and data analysis community
\cite{eelevenlivretheodoridistheory}.
The more challenging
version of
that problem is the
blind
\cite{a593}
or unsupervised one,
where the input values are unknown 
and uncontrolled,
but it may be known that the
input signal belongs to a given
class
(due to this partial knowledge, these methods are sometimes stated to be
semi-blind).
Both versions may then be extended 
to
multiple-input multiple-output (MIMO) systems.

Besides, in various applications, what is needed is not
the direct transform performed by the above system, but the inverse
of that transform (assuming it is invertible).
For SISO non-blind and  blind configurations,
this is motivated by the fact that one eventually
only accesses the ouput
\yssmixsigdisctimecentvec\
of the above direct system, and one aims at deriving a signal
\yssoutsepsystsigdisctimecentvec\
which ideally restores the original
source signal
\ysssrcsigdisctimecentvec .
To this end, one may 
\ymodifartitionehundredsixvonestepone{first}
use the above-mentioned 
system identification methods in order to 
\ymodifartitionehundredsixvonestepone{estimate} the direct system,
then derive its inverse
and eventually transfer the 
output 
\yssmixsigdisctimecentvec\
of the direct system through the inverse system.
Alternatively, one may develop
methods for initially
identifying the \emph{inverse} system itself.
Extended versions 
of
this 
``(unknown)
system inversion'' task
concern MIMO configurations,
where a set of original source signals
\ysssrcsigdisctimecentone\ to
\ysssrcsigdisctimecentlast\
are to be respectively 
restored on
the outputs
\yssoutsepsystsigdisctimecentone\
to
$
\yssoutsepsystsigdisctimecentvec
_
\Ysssrcnb
$
of the inverse system.

The blind MIMO version of the above system inversion problem is
almost the same as
blind source separation (BSS) 
\cite{book-comon-jutten-ap},\cite{amoi6-48},\cite{icabook-oja}:
as in 
system inversion,
BSS 
aims at canceling 
the contributions of
all sources but one in each output signal of the separating 
system;
however, in BSS, one often 
allows each output
signal to be equal
to a source signal only up to an
acceptable
residual transform.
These transforms,
called indeterminacies,
cannot be avoided because only limited constraints
are set on the 
source signals 
and on the direct system 
which combines
(i.e., ``mixes'', in BSS terms) 
these signals.
In particular,
the first class of BSS methods that was developed and that is still of
major importance is Independent Component Analysis, or ICA
\cite{book-comon-jutten-ap},\cite{amoi6-48},\cite{icabook-oja},
which may be seen as an extension of more conventional Principal
Component Analysis, or PCA
\cite{book-jolliffe-pca}
(PCA alone cannot achieve
BSS
\cite{icabook-oja}).
ICA is a statistical approach, which essentially
requires statistically independent random source signals.
Thus,
for the simplest class of mixtures, ICA is guaranteed to restore
the source signals up to limited indeterminacies
\cite{book-comon-jutten-ap},\cite{amoi6-48},\cite{icabook-oja}.

We now consider
quantum information processing (QIP)
\cite{booknielsen}
and quantum machine learning
\cite{amq73},\cite{amq72bis},\cite{amq74},
i.e.
processing of quantum data and/or processing with quantum means,
and we still focus on system identification and inversion problems.
Among these problems,
the one 
which was
first
studied is the quantum version of non-blind 
system identification,
especially
\ymodifartitionehundredsixvonestepone{\cite{footnote-early-references}}
introduced in 1997 in 
\cite{amq30official}
and
called
``quantum process tomography'' or QPT
by the QIP 
community: see
e.g.
\cite{amq-baldwin-physreva-2014},
\cite{amq75},
\cite{amq45},
\cite{amq50-physical-review},
\cite{amq59},
\cite{booknielsen},
\cite{amq48},
\cite{amq52-physical-review},
\cite{amq56},
\cite{paper-white-gilchrist-2007}.
The quantum version of the above-mentioned classical
source separation,
called Quantum Source Separation, or QSS,
and especially its blind version, or BQSS, were then introduced in
2007 in
\cite{amoi5-31}.
Two main classes of BQSS methods were developed since then.
The first one may be seen as a quantum extension of the above-mentioned
classical ICA methods, since it
takes advantage of the statistical independence of
the parameters that define random source quantum states
(qubit states).
It is called Quantum Independent Component Analysis
(or QICA, see e.g.
\cite{amoi5-31},\cite{amoi6-18})
or, more precisely,
Quantum-Source Independent Component Analysis
(or QSICA, see e.g.
\cite{amoi6-42})
to insist on the quantum nature of the considered source data,
whereas it uses classical processing means
(after quantum/classical data conversion).
The second main class of BQSS methods was introduced in 2013-2014 in
\cite{amoi6-34},\cite{amoi6-37}
and then especially detailed in
\cite{amoi6-64}.
It is based on the unentanglement of the considered source quantum states
and it typically
uses quantum processing means to restore these unknown states
from their coupled version.
Independently from the above quantum extensions of ICA, a quantum version
of PCA was introduced in 2014 in
\cite{amq72}.

In the present paper, our first contribution 
(see Section \ref{sec-method-multiple-prep})
concerns
yet another type of unsupervised quantum
machine learning methods
\cite{amq73}, namely Blind Quantum Process Tomography,
or BQPT.
Here again, the term ``blind'', or ``unsupervised'',
refers to the fact that we consider
situations where the input values of the process to be identified
are unknown, 
but they are requested to meet some 
(hereafter statistical) properties.
We briefly introduced that BQPT concept in 2015 in
\cite{amoi6-46}
and we only
outlined some
resulting BQPT methods in that and some subsequent short conference
papers,
but only as spin-offs of corresponding BQSS methods.
On the contrary, the present paper is the first one where we provide
a detailed description of a method which combines the following
features:
\begin{enumerate}
\item This method is primarily intended for BQPT, not for BQSS.
To this end, it only uses classical processing means: on the contrary, 
using quantum processing means requires one to precisely
characterize them beforehand, which is a significant
drawback here, since
BQPT, as QPT, is especially developed as a tool for characterizing
quantum gates, as discussed in Sections
\ref{sec-estimate-process}
and
\ref{sec-conclusion}.
\item This BQPT method therefore first performs measurements at the output of
the system, i.e. quantum process, to be identified
(see Section \ref{sec-measurements}), in order to convert 
quantum states into classical-form data before they are processed
with classical means.
\item Moreover, for the Heisenberg coupling process considered below
as an example (see Section 
\ref{sec-heisenberg process}), we aim at minimizing the number
of types of measurements performed to fully characterize that process.
\end{enumerate}

As detailed further in this paper,
usual, i.e. non-blind, QPT, as well as the above first form of BQPT,
use sample frequencies
of the above-mentioned measurement outcomes at the
output of the process
(i.e. 
normalized cumulative values associated with these outcomes), 
derived from many copies of
each considered state value.
To apply such methods, one should therefore
be able to \emph{prepare} many copies of the \emph{same} 
input state, which is cumbersome.
Our second contribution in this paper 
(see Section \ref{sec-method-single-prep})
then consists of extensions of the above BQPT methods, which
also allow one to use few copies or even one instance
(i.e. preparation) of each quantum state.
The numerical performance of this second type of methods is reported in
Section \ref{sec-tests},
whereas its applications 
are presented in Section \ref{sec-conclusion},
together with conclusions drawn from this investigation.

\section{Considered quantum process and
state properties}\label{sec-heisenberg process}
In QPT and BQPT problems, the system or
process to be identified receives a quantum
state
\ymixsyststateinitial ,
e.g. associated with a set of qubits considered at time
\yqubitonetimeinit .
This system outputs a quantum state
\ymixsyststatefinal\
associated with the same set of qubits at a later time
\yqubitonetimefinal .
The behavior of the system is defined by an operator which may be rather general
(e.g. an arbitrary unitary operator) or which may be restricted to
a given class of operators corresponding to a specific type of physical
devices (see e.g. \cite{amq59}).
In this paper, we address the second case and we
consider a device composed of two distinguishable 
qubits
\cite{amoi6-64}
implemented as electron spins 1/2, that are coupled according to the
cylindrical-symmetry Heisenberg model,
which is e.g. relevant for spintronics applications.
We stress that this type of
coupling is only used as a concrete example, to show
how to fully implement the proposed concepts in a relevant case, but that
these concepts and resulting practical BQPT algorithms may then
be transposed by the reader to other classes of quantum processes
and associated applications.

The symmetry axis of the Heisenberg model is here denoted as $Oz$. 
The considered
spins are supposed to be placed in a magnetic
field (also oriented
along $Oz$ and with a magnitude \ymagfieldnot) and thus coupled to it.
Moreover, we assume an isotropic
$\overline{\overline{g}}$
tensor,
with principal value $g$.
The time interval when these spins are
considered
is 
supposed to be short
enough for 
their coupling with
their environment 
to be
negligible.
In these conditions, 
the temporal evolution of the state of
the device composed of these two spins
is governed by the following
Hamiltonian:
\yeqdefhamiltonian
where:
\begin{list}{}{\setlength{\leftmargin}{10mm}
	       \setlength{\labelwidth}{8mm}
	       \setlength{\labelsep}{2mm}}
\item[$\bullet$]
$\Yhamiltonfieldscale = g \mu_e$, where
$\mu_e$ is the Bohr magneton, i.e. $\mu_e = e \hbar / 2 m_e =
0.927 \times 10^{-23} J T^{-1}$
and
$
\hbar
$
is the reduced Planck constant,
\item[$\bullet$]
$s_{ix}, \ s_{iy}, \ s_{iz}$, with 
$
\Yqubitoneindexstd
\in
\{
1,
2
\},
$
are the three components of the vector operator 
\overrightarrow{s_{i}}
associated 
with
spin $i$ 
in a cartesian frame,
\item[$\bullet$]
$J_{xy}$ and $J_{z}$ are the principal values of the exchange tensor.
\end{list}
Among the above parameters,
the value of $g$
may be experimentally determined,
and
\ymagfieldnot\
can be 
measured.
The values of
$J_{xy}$ and $J_{z}$ 
are 
here assumed to be
unknown.

We 
here suppose that 
each spin
\yqubitoneindexstd ,
with
$
\Yqubitoneindexstd
\in
\{
1,
2
\},
$
is prepared, i.e. initialized,
at a given time
\yqubitonetimeinit,
\ymodifartitionehundredsixvonestepone{in}
the pure
state
\yeqdeftwoqubitstateinitindexi
where
$
|+ \rangle
$
and
$
|- \rangle
$
are
eigenkets of
$s_{iz}$,
for the eigenvalues
$1/2$ and $- 1/2$ respectively.
We will further use the polar representation of the
qubit parameters
$\alpha_i$
and
$\beta_i$,
which reads
\yeqdefqubitpolarqubitindexstd
where \ysqrtminusone\ is the imaginary unit, and
with
$
0
\leq
\Yparamqubitindexstdstateplusmodulus
\leq
1
$
and
\yeqdefparamqubitindexstdstateminusmodulusvsparamqubitindexstdstateplusmodulus
because
each spin state
$| \psi_i 
(
\Yqubitonetimeinit
)
\rangle$
has unit norm.
Moreover, for each couple of phase parameters
\yparamqubitindexstdstateplusphase\
and
\yparamqubitindexstdstateminusphase ,
only their difference has a physical meaning.
After they have been prepared,
these
spins are coupled 
according to
the above-defined
model for
$
t
\geq
\Yqubitonetimeinit
$.

Hereafter,
we consider the state
of the overall system composed of these two
distinguishable
spins.
At time
\yqubitonetimeinit, this state is equal to the tensor product of the
states
of both spins
defined in
(\ref{eq-twoqubit-state-init-index-i}).
It therefore reads
\yeqdefqubitbothtimeinitstatetensorprod
in the four-dimensional basis
$
\Ytwoqubitsbasisplusplus
=
\{ | ++ 
\rangle
, | +- 
\rangle
, | -+ 
\rangle
, | -- 
\rangle
\}
$.

The state of this two-spin system then evolves with time.
Its value 
\ymixsyststatefinal\
at any subsequent time $t$
may be derived from its above-defined Hamiltonian. 
\ymodifartitionehundredsixvonestepone{It is defined
\cite{amoi6-18}} by
\yeqdefstatetfinalvsopmixstatetinitcomponents
where
\yveccompsyststatetinit\
and
\yveccompsyststatetfinal\
are the column vectors of components of
\ymixsyststateinitial\ and
\ymixsyststatefinal , respectively,
in basis
\ytwoqubitsbasisplusplus 
. For instance,
as shown by
(\ref{eq-etat-deuxspin-plusplus-initial-decompos}),
\yeqdefveccompsyststatetinit
where 
$
^T
$ 
stands for transpose.
Moreover, the matrix
\yopmix\ of 
(\ref{eq-statetfinalvsopmixstatetinit-components}),
which defines
the transform
applied to
\ymixsyststateinitial,
reads
\yeqdefopmixmatrixdecompose
with
\yeqdefopmixbasesdef
and
\ymodifartitionehundredsixvonestepone{\yopmixdiag\
equal to} 
\yeqdefopmixdiagdef
The four real (angular) frequencies
$
\omega _{1 , 1}
$
to
$
\omega _{1 , -1}
$
in
(\ref{eq-opmixdiagdef})
depend on 
the
physical setup.
In
\cite{amoi6-18},
it was
shown that they read
\yeqdefomegaallexpress
Since
the values of the parameters
$J_{xy}$ and $J_{z}$
of the Hamiltonian of 
(\ref{eq-online-hamiltonian})
are 
presently unknown,
the values of the parameters
$
\omega _{1 , 1}
$
to
$
\omega _{1 , -1}
$
of the quantum process involved in
(\ref{eq-statetfinalvsopmixstatetinit-components})
are also unknown.

In this paper, we address the (B)QPT problem, i.e. we aim at estimating
the matrix
\yopmix\
involved in
(\ref{eq-statetfinalvsopmixstatetinit-components}),
which defines the considered quantum process.
Moreover, we estimate it in a blind, i.e. unsupervised, way, that is:
\begin{itemize}
\item
by using values of the output state
\ymixsyststatefinal\
of this process,
\item
without using nor knowing values of its input state
\ymixsyststateinitial,
\item
but by knowing and exploiting some properties of
these states
\ymixsyststateinitial.
In this paper,
these requested properties are as follows.
The states
\ymixsyststateinitial\
are required to be unentangled (as shown by
(\ref{eq-qubitbothtimeinitstate-tensor-prod})).
Besides, the proposed BQPT methods
are statistical approaches and the six parameters
\yparamqubitindexstdstateplusmodulus,
\yparamqubitindexstdstateplusphase\
and
\yparamqubitindexstdstateminusphase,
with
$\Yqubitindexstd \in \{ 1 , 2 \}$,
defined in
(\ref{eq-def-qubit-polar-qubit-indexstd})
are constrained to have
properties that are similar
to those requested in the above-mentioned QSICA methods:
(i) these parameters are random valued, so that we here consider
random pure quantum states
$| \psi_i 
(
\Yqubitonetimeinit
)
\rangle$
(see 
\cite{amoi6-67}
for more details)
and
(ii) some combinations of
the random variables (RVs)
\yparamqubitindexstdstateplusmodulus,
\yparamqubitindexstdstateplusphase\
and
\yparamqubitindexstdstateminusphase\
are statistically independent
and have a few known statistical features,
as detailed further in this paper.
\end{itemize}

As explained in Section
\ref{sec-intro},
the considered BQPT task is performed by using only
classical-form processing means.
To this end, the available quantum-form data, namely the
output states
\ymixsyststatefinal,
are first converted into classical-form data, by means of
measurements, as described hereafter.

\section{Measurements for process outputs}\label{sec-measurements}
The first type of proposed BQPT approaches uses a set of
copies of each output state
\ymixsyststatefinal.
For each copy, it measures the components of the considered two spins
along the above-defined $Oz$ direction.
The result of
each such measurement has
four possible values,
that is
$(+\frac{1}{2},+\frac{1}{2})$,
$(+\frac{1}{2},-\frac{1}{2})$,
$(-\frac{1}{2},+\frac{1}{2} )$
or
$(-\frac{1}{2},-\frac{1}{2})$
in normalized units
(see Appendix \ref{sec-stochastic-QIP-standard-concepts}).
Their probabilities are respectively denoted as
\ytwoqubitsprobaplusplusdirzz\
to
\ytwoqubitsprobaminusminusdirzz\
hereafter.
Using the polar representation
(\ref{eq-def-qubit-polar-qubit-indexstd}),
these
probabilities read
\cite{amoi6-18},\cite{amoi6-42}
\yeqdefstatecoefvstwoqubitsprobathreepolar
with
\yeqdefeqdeftwoqubitresultphaseinittwoqubitresultphaseevol
\ymodifartitionehundredsixvonestepone{Probability
\ytwoqubitsprobaminusplusdirzz\
is not considered hereafter because
the sum of
\ytwoqubitsprobaplusplusdirzz\
to
\ytwoqubitsprobaminusminusdirzz\
is equal to 1.}

In practice, for each value of state
\ymixsyststatefinal,
estimates of probabilities
\ytwoqubitsprobaplusplusdirzz\
to
\ytwoqubitsprobaminusminusdirzz\
are derived, typically as the sample frequencies of
the associated measurement outcomes obtained for all
copies of
\ymixsyststatefinal\
(see e.g.
\cite{amq75},\cite{amq30official},\cite{amoi6-18},\cite{amoi6-42}).

Similarly, these BQPT approaches use another set of
copies of each output state 
\ymixsyststatefinal,
by measuring the two spin components along an axis $Ox$
which is orthogonal to $Oz$.
These measurements
yield the same four possible outcomes as above,
but with different
probabilities, which
are denoted as
\ytwoqubitsprobaplusplusdirxx\
to
\ytwoqubitsprobaminusminusdirxx\
hereafter.
As shown in
\cite{amoi6-99},
these probabilities  have the following properties
\yeqdeftwoqubitsprobaplusplusdirxxminustwoqubitsprobaminusminusdirxxdefandexplicit
where
\yeqdeftwoqubitsprobadirxxplusplusdiffminusminusphasediffvsexchangetensor
The value of
\ytwoqubitsprobadirxxplusplussumminusminusphasediff\
in
(\ref{eq-combine-omegaoneone-omegaoneminusone})
is known, since it can be derived from the above-defined
known quantities.
Moreover
\yeqdeftwoqubitsprobaplusplusdirxxplustwoqubitsprobaminusminusdirxx

The BQPT methods proposed in this paper therefore consist of two
major
steps. The first step aims at estimating the 
unknown values of the parameters
\ytwoqubitresultphaseevolsin,
\ytwoqubitsprobadirxxplusplusdiffminusminusphasediffcosgen\
and
\ytwoqubitsprobadirxxplusplusdiffminusminusphasediffsingen\
of the mappings from 
the parameters
\yparamqubitindexstdstateplusmodulus,
\yparamqubitindexstdstateplusphase\
and
\yparamqubitindexstdstateminusphase\
of the initial qubit states
\ymixsyststateinitial\
to the probabilities
of measurement outcomes,
namely
\ytwoqubitsprobaindexstddirzz\
and
\ytwoqubitsprobaindexstddirxx ,
with
$\Yqubitbothtimefinalstatecoefindexindexstd =1$ to 4,
or their combinations.
The second step then uses the estimated values of
\ytwoqubitresultphaseevolsin,
\ytwoqubitsprobadirxxplusplusdiffminusminusphasediffcosgen\
and
\ytwoqubitsprobadirxxplusplusdiffminusminusphasediffsingen\
to derive an estimate of matrix
\yopmixdiag\
of
(\ref{eq-opmixmatrixdecompose})
and hence of the complete matrix
\yopmix\
of 
(\ref{eq-opmixmatrixdecompose}),
which defines the considered process.
We now proceed to the description of these methods.

\section{Multiple-preparation BQPT methods}\label{sec-method-multiple-prep}
We first consider the estimation of parameter
\ytwoqubitresultphaseevolsin.
This is achieved by exploiting
(\ref{eq-statecoef-vs-twoqubitsprobaplusminus-polar-versionthree}).
If we were developing a conventional, i.e. non-blind, QPT method,
we would use one or several instances of Eq.
(\ref{eq-statecoef-vs-twoqubitsprobaplusminus-polar-versionthree}),
and each of these instances would involve (i) known values of the
input parameters
\yparamqubitindexstdstateplusmodulus,
\yparamqubitindexstdstateplusphase,
\yparamqubitindexstdstateminusphase\
and hence
\ytwoqubitresultphaseinit\
of
\ymixsyststateinitial\
and (ii)
an estimate of the set of output probabilities
\ytwoqubitsprobaindexstddirzz,
derived from a set of copies of
\ymixsyststatefinal.
This approach is constraining because it requires one to
\emph{precisely} prepare each state \emph{value}
$| \psi_i 
(
\Yqubitonetimeinit
)
\rangle$
(for state preparation and associated errors, see e.g.
\cite{amq30official}),
otherwise the errors in
\yparamqubitindexstdstateplusmodulus\
and
\ytwoqubitresultphaseinit\
yield errors in the 
estimated value of
\ytwoqubitresultphaseevolsin.

The above drawback is avoided by our BQPT methods.
Following the above-defined terminology, 
these methods are blind in the sense that
they estimate
\ytwoqubitresultphaseevolsin\
by
using only 
a set 
of 
estimated 
values of
output probabilities
\ytwoqubitsprobaindexstddirzz,
without knowing the values of the 
input parameters
\yparamqubitindexstdstateplusmodulus,
\yparamqubitindexstdstateplusphase,
\yparamqubitindexstdstateminusphase\
and hence
\ytwoqubitresultphaseinit,
but requesting them to 
have some known properties.
More precisely, we here consider statistical methods, which
operate
with a set of
random states
$| \psi_i 
(
\Yqubitonetimeinit
)
\rangle$
and which thus only set constraints on some of the
\emph{statistical parameters} of
(combinations of)
\yparamqubitindexstdstateplusmodulus,
\yparamqubitindexstdstateplusphase,
\yparamqubitindexstdstateminusphase ,
not on their \emph{individual values} 
for each state
$| \psi_i 
(
\Yqubitonetimeinit
)
\rangle$.
In particular, the versions of these BQPT methods
considered in this paper use only the first-order mean statistics
of the available quantities 
\ytwoqubitsprobaindexstddirzz,
i.e. their expectations
$E \{ 
\Ytwoqubitsprobaindexstddirzz \}$.
When assuming
\yparamqubitonestateplusmodulus,
\yparamqubittwostateplusmodulus\
and
\ytwoqubitresultphaseinit\
to be statistically independent RVs,
(\ref{eq-statecoef-vs-twoqubitsprobaplusminus-polar-versionthree})
yields
\yeqdefstatecoefvstwoqubitsprobaplusminuspolarversionthreeexpect
In this equation,
$E \{ 
\Ytwoqubitsprobaplusminusdirzz \}$
is known:
in practice, it is
estimated as the sample mean of the estimates of
all values of \ytwoqubitsprobaplusminusdirzz, themselves typically
estimated
with sample frequencies, as explained above.
Besides, as detailed e.g. in
\cite{amoi6-18},\cite{amoi6-42}
for BQSS methods intended for the Heisenberg coupling model, 
setting the constraint
\yeqdefparamqubitonestateplusmodulusltonehalfltparamqubittwostateplusmodulus
allows one to derive
\yparamqubitonestateplusmodulus\
and
\yparamqubittwostateplusmodulus\
from
(\ref{eq-statecoef-vs-twoqubitsprobaplusplus-polar})
and
(\ref{eq-statecoef-vs-twoqubitsprobaminusminus-polar-vs-paramqubitindexstdstateplusmodulus})
without any \ymodifartitionehundredsixvonestepone{ambiguity}, for each unknown state
\ymixsyststateinitial.
This yields
\yeqdefqubitindexstdstateplusmodulussolbothfrommlsp
with
$\epsilon
_{1}
=
- 1$
and
$\epsilon
_{2}
=
1$.
Taking the sample mean of any function of
\yparamqubitindexstdstateplusmodulus\
defined by
(\ref{eq-qubitindexstdstateplusmodulus-sol-both-from-mlsp})
then yields estimates of all statistics of
\yparamqubitindexstdstateplusmodulus\
involved in
(\ref{eq-statecoef-vs-twoqubitsprobaplusminus-polar-versionthree-expect}).
Finally, we only set the following constraint on one statistical
parameter of \ymodifartitionehundredsixvonestepone{the} used values of 
\yparamqubitindexstdstateplusphase\
and
\yparamqubitindexstdstateminusphase,
again without having to know their individual values.
We request the states 
\ymixsyststateinitial\
to be prepared with a procedure which is such that the value of
$
E
\{
\sin \Ytwoqubitresultphaseinit
\}
$
is known.
With these constraints on input state statistics, the only unknown in
(\ref{eq-statecoef-vs-twoqubitsprobaplusminus-polar-versionthree-expect})
is
\ytwoqubitresultphaseevolsin.
By solving this type of equations,
this BQPT method then yields the desired estimate of
\ytwoqubitresultphaseevolsin.
In particular,
a simple case consists of
using
$E
\{
\sin \Ytwoqubitresultphaseinit
\}
=
0$
(which may e.g. be achieved by preparing the two spins
with states such that
$(
\Yparamqubitonestateminusphase
-
\Yparamqubitonestateplusphase )$
and
$( \Yparamqubittwostateminusphase
-
\Yparamqubittwostateplusphase
)$
are statisticially independent and
have the same statistics):
then,
(\ref{eq-statecoef-vs-twoqubitsprobaplusminus-polar-versionthree-expect})
straightforwardly yields
\yeqdeftwoqubitresultphaseevolsinvsfirstordermomentsqr
In some configurations
the sign of
\ytwoqubitresultphaseevolsin\
is known
\cite{amoi6-18},\cite{amoi6-42},
so that 
the
value of 
\ytwoqubitresultphaseevolsin\
may be derived
from
(\ref{eq-twoqubitresultphaseevolsin-vs-first-order-moment-sqr}).
Otherwise,
it may be derived from another instance of
(\ref{eq-statecoef-vs-twoqubitsprobaplusminus-polar-versionthree-expect}),
using data that yield another value of
$E
\{
\sin \Ytwoqubitresultphaseinit
\}$:
details about how to solve this sign indeterminacy
and how to
also estimate parameters
\ytwoqubitsprobadirxxplusplusdiffminusminusphasediffcosgen\
and
\ytwoqubitsprobadirxxplusplusdiffminusminusphasediffsingen\
are provided below for an improved version of our methods.
Indeed, the above version of BQPT 
is attractive because it does not require each value of
\ymixsyststateinitial\
to be known, but it
still yields a limitation:
it requires one to be able to prepare the
\emph{same} value 
\ymixsyststateinitial\
a large number of times, to derive an associated
frequency-based estimate of each set of probabilities
\ytwoqubitsprobaindexstddirzz.
This still requires some control of the input states of the process,
that we would like to avoid, 
in order to \ymodifartitionehundredsixvonestepone{simplify}
the practical operation of BQPT
methods and
to make them ``\ymodifartitionehundredsixvonestepone{blinder}''.
We hereafter show how to avoid this preparation of many copies of
each state
\ymixsyststateinitial .

\section{Single-preparation BQPT methods}\label{sec-method-single-prep}

\subsection{Single-preparation QIP}\label{sec-method-single-prep-principle-prep}
As a second contribution in this paper, we now extend (B)QPT
methods so that they can operate 
with a few copies or even a single instance of each considered
input state
\ymixsyststateinitial.
For non-blind methods as defined above, this does not seem to be possible,
because they need many copies of each state
\ymixsyststateinitial\
and associated outcomes of measurements performed for each state
\ymixsyststatefinal,
in order to derive a frequency-based estimate of each set of
probabilities
\ytwoqubitsprobaindexstddirzz.
On the contrary, our blind versions of QPT can be extended so as
to reach this goal, because they only need one to estimate
\emph{expectations} of these (now random) probabilities
\ytwoqubitsprobaindexstddirzz,
i.e.
$E \{ 
\Ytwoqubitsprobaindexstddirzz \}$,
not each of their individual values 
\ytwoqubitsprobaindexstddirzz\
for each state
\ymixsyststateinitial .
In the short conference paper 
\cite{amoi6-104},
we very recently introduced a general QIP framework
(i.e., not restricted to BQPT) for estimating
expectations
$E \{
\Ytwoqubitsprobaindexstd
\}$
of probabilities
\ytwoqubitsprobaindexstd\
of outcomes of general types of quantum measurements.
Its principle is summarized hereafter,
whereas its detailed description and properties are
provided in Appendix~\ref{sec-stochastic-QIP}.

For each expectation
$E \{
\Ytwoqubitsprobaindexstd
\}$
of a random probability
\ytwoqubitsprobaindexstd\ 
to be estimated, as discussed
above, in practice the expectation operator
$E \{ . \}$
is replaced by a sample mean, i.e. by a \emph{sum} (of values,
moreover normalized). Similarly, each probability
\ytwoqubitsprobaindexstd\ 
is replaced by a sample frequency, i.e. by a \emph{sum} (of 1 and
0, depending whether the considered event occurs or not for
each trial defined by a preparation of the initial quantum
states 
(\ref{eq-twoqubit-state-init-index-i})
and by
an associated measurement of a couple of spin components;
this summation is
here again followed
by a normalization, by the total number of trials).
$E \{
\Ytwoqubitsprobaindexstd
\}$
is therefore estimated by a 
(normalized)
``sum of sums'', 
which may then be reinterpreted as a single
global
sum, and what primarily matters is the total number of 
preparations of initial quantum
states 
(\ref{eq-twoqubit-state-init-index-i})
involved in that global sum,
whereas the number of preparations for each state value
(\ref{eq-twoqubit-state-init-index-i})
may be decreased, down to 1,
as
\ymodifartitionehundredsixvonestepone{confirmed by simulations}
in Section
\ref{sec-tests}
(which also justifies why even better performance is obtained
when decreasing the number of preparations per state for a 
given total number of preparations,
i.e. while increasing accordingly the considered number
of different states).
The corresponding BQPT methods are therefore called
\ymodifartitionehundredsixvonestepone{single-preparation}
BQPT
methods.
It should be clear that
they can be freely used with
either one instance or several (e.g. many) copies per state, i.e. 
the above terminology means that
these methods
\emph{allow} one to use a single instance of each state.
On the contrary, our so-called 
\ymodifartitionehundredsixvonestepone{multiple-preparation}
BQPT
methods \emph{force} one to use many state copies to achieve
good performance.

\subsection{Estimating the parameter of $Oz$ measurements}\label{sec-estim-v}
We here aim at using the single-preparation approach of Section
\ref{sec-method-single-prep-principle-prep} to estimate the parameter
\ytwoqubitresultphaseevolsin\ involved in the probabilities
\ytwoqubitsprobaindexstddirzz.
We hereafter again take advantage of
(\ref{eq-statecoef-vs-twoqubitsprobaplusminus-polar-versionthree-expect}),
and especially of its version
(\ref{eq-twoqubitresultphaseevolsin-vs-first-order-moment-sqr}),
derived in Section \ref{sec-method-multiple-prep}
for our multiple-preparation BQPT method.
However,
these expressions involve
$E \{ \Yparamqubitonestateplusmodulus ^2 \}$
and
$E \{ \Yparamqubittwostateplusmodulus ^2 \}$
which,
unlike in
Section \ref{sec-method-multiple-prep},
cannot here be estimated by using
the expectation of the square of
(\ref{eq-qubitindexstdstateplusmodulus-sol-both-from-mlsp})
because
this involves
expectations of
\emph{nonlinear combinations}
of the above probabilities
\ytwoqubitsprobaplusplusdirzz\
to
\ytwoqubitsprobaminusminusdirzz,
whereas we here aim at developing a 
\emph{single-preparation} algorithm,
for which 
Section
\ref{sec-method-single-prep-principle-prep}
only defined how to estimate the 
expectations
of
\ytwoqubitsprobaplusplusdirzz\
to
\ytwoqubitsprobaminusminusdirzz\
themselves.
We here solve this problem by using a modified approach, where
we first take the expectation of
(\ref{eq-statecoef-vs-twoqubitsprobaplusplus-polar})
and
(\ref{eq-statecoef-vs-twoqubitsprobaminusminus-polar-vs-paramqubitindexstdstateplusmodulus}),
again
for statistically independent RVs
\yparamqubitonestateplusmodulus\
and
\yparamqubittwostateplusmodulus.
This yields
\yeqdefstatecoefvstwoqubitsprobathreepolarmean
These equations involve only the unknown of interest,
$E \{ \Yparamqubitonestateplusmodulus ^2 \}$
and
$E \{ \Yparamqubittwostateplusmodulus ^2 \}$.
Again setting the constraint
(\ref{eq-qubit-modulus-allowed-range}),
they yield the unique solution
\yeqdefparamqubitindexstdstateplusmodulussolbothdirzz
again with
$\epsilon
_{1}
=
- 1$
and
$\epsilon
_{2}
=
1$.
If
the sign of
\ytwoqubitresultphaseevolsin\
is known,
the
value of 
\ytwoqubitresultphaseevolsin\
may thus be derived
from
(\ref{eq-twoqubitresultphaseevolsin-vs-first-order-moment-sqr}),
therefore using data such that
$
E
\{
\sin \Ytwoqubitresultphaseinit
\}
=
0
$.
Otherwise,
(\ref{eq-twoqubitresultphaseevolsin-vs-first-order-moment-sqr})
is first used to estimate
$
\Ytwoqubitresultphaseevolsin
^
2
$,
which yields
$
|
\Ytwoqubitresultphaseevolsin
|
$,
and the sign of
\ytwoqubitresultphaseevolsin\
is
then derived from another 
set of spin state preparations,
now
considering the case when
$E
\{
\sin \Ytwoqubitresultphaseinit
\}
\neq
0$.
Eq.
(\ref{eq-statecoef-vs-twoqubitsprobaplusminus-polar-versionthree-expect})
then yields
\yeqdeftwoqubitresultphaseevolsinvsitssqr
Taking the sign of this equation, where a factor is guaranteed to
be positive, results in
\yeqdeftwoqubitresultphaseevolsinvsitssqrsgn
For this second set of spin state preparations, 
(i) we do not request the
\emph{value} of
$E
\{
\sin \Ytwoqubitresultphaseinit
\}$
but only its \emph{sign}
to be known,
(ii)
the
values of
$E
\{
\Yparamqubitonestateplusmodulus ^2 
\}$,
$E
\{
\Yparamqubittwostateplusmodulus ^2 
\}$
and
$E
\{
\Ytwoqubitsprobaplusminusdirzz
\}$
may again be estimated as explained above.
Also using the above estimate of
$\Ytwoqubitresultphaseevolsin
^2$,
Eq.
(\ref{eq-twoqubitresultphaseevolsin-vs-its-sqr-sgn})
then allows one to estimate the sign of
\ytwoqubitresultphaseevolsin.

\subsection{Estimating the parameters of $Ox$ measurements}\label{sec-estim-w-both}
We then show how to estimate the parameters
\ytwoqubitsprobadirxxplusplusdiffminusminusphasediffcosgen\
and
\ytwoqubitsprobadirxxplusplusdiffminusminusphasediffsingen\
of
(\ref{eq-twoqubitsprobaplusplusdirxx-minus-twoqubitsprobaminusminusdirxx-def-and-explicit}),
using measurements of spin components along the $Ox$ axis, in addition to
the $Oz$
axis, and the corresponding expectations
$E \{ 
\Ytwoqubitsprobaindexstddirxx \}$
and
$E \{ 
\Ytwoqubitsprobaindexstddirzz \}$.
Here again, we only constrain the statistical parameters of the
RVs
\yparamqubitindexstdstateplusmodulus\
and
$
(
\Yparamqubitindexstdstateminusphase
-
\Yparamqubitindexstdstateplusphase
)
$,
not their individual deterministic values, in order to be able to solve
(\ref{eq-twoqubitsprobaplusplusdirxx-minus-twoqubitsprobaminusminusdirxx-def-and-explicit})
with respect to
\ytwoqubitsprobadirxxplusplusdiffminusminusphasediffcosgen\
and
\ytwoqubitsprobadirxxplusplusdiffminusminusphasediffsingen .
More precisely,
the RVs
\yparamqubitonestateplusmodulus , 
\yparamqubittwostateplusmodulus ,
$(
\Yparamqubitonestateminusphase
-
\Yparamqubitonestateplusphase )$
and
$( \Yparamqubittwostateminusphase
-
\Yparamqubittwostateplusphase
)$
are here statistically independent.
Besides,
\yparamqubitonestateplusmodulus\
and
\yparamqubittwostateplusmodulus\
have the same statistics.
Finally,
$(
\Yparamqubitonestateminusphase
-
\Yparamqubitonestateplusphase )$
and
$( \Yparamqubittwostateminusphase
-
\Yparamqubittwostateplusphase
)$
have the same statistics, moreover with
\yeqdefparamqubitindexstdstateplusandminusphaseconstraints
which is e.g. obtained with RVs
$(
\Yparamqubitindexstdstateminusphase
-
\Yparamqubitindexstdstateplusphase
)$
whose probability density functions are
even 
and non-zero on
$
[
-
\frac{\pi}{2}
,
\frac{\pi}{2}
]
$.
In that case,
(\ref{eq-statecoef-vs-twoqubitsprobaplusplus-polar}),
(\ref{eq-twoqubitsprobaplusplusdirxx-minus-twoqubitsprobaminusminusdirxx-def-and-explicit})-%
(\ref{eq-twoqubitsprobadirxxplusplusdiffminusminusfactorimag-explicit})
and
(\ref{eq-twoqubitsprobaplusplusdirxx-plus-twoqubitsprobaminusminusdirxx-explicit})
yield
\ymodifartitionehundredsixvonestepone{\cite{footnote-no-constraint-rone-rtwo}}
(with the same statistics for
$\Yqubitindexstd
=
1$
and
2):
\yeqdeftwoqubitsprobaplusplusdirzzmeanandothers
Once
$E
\{
\Ytwoqubitsprobaplusplusdirzz
\}$,
$E
\{
\Ytwoqubitsprobaplusplusdirxx
\}$
and
$E
\{
\Ytwoqubitsprobaminusminusdirxx
\}$
have been estimated
as explained above,
Eq.
(\ref{eq-twoqubitsprobaplusplusdirzz-mean}),
with
$E
\{
\Yparamqubitindexstdstateplusmodulus
^2
\}
\geq
0$
due to
$\Yparamqubitindexstdstateplusmodulus
\geq
0$,
yields
\yeqdeftwoqubitsprobaplusplusdirzzmeansqrt
Moreover,
$\Yparamqubitindexstdstateplusmodulus
\geq
0$,
(\ref{eq-cos-paramqubitindexstdstateminusminusplusphase-positive})
and
(\ref{eq-twoqubitsprobaplusplusdirxx-plus-twoqubitsprobaminusminusdirxx-mean})
yield
\yeqdeftwoqubitsprobaplusplusdirxxplustwoqubitsprobaminusminusdirxxmeanresultingstat
Using
(\ref{eq-combine-omegaoneone-omegaoneminusone}),
(\ref{eq-twoqubitsprobaplusplusdirzz-mean-sqrt})
and
(\ref{eq-twoqubitsprobaplusplusdirxx-plus-twoqubitsprobaminusminusdirxx-mean-resulting-stat}),
Eq.
(\ref{eq-twoqubitsprobadirxxplusplusdiffminusminusfactorreal-explicit-mean})
and
(\ref{eq-twoqubitsprobadirxxplusplusdiffminusminusfactorimag-explicit-mean})
then
yield estimates of
$E
\{
\Ytwoqubitsprobadirxxplusplusdiffminusminusfactorreal
\}$
and
$E
\{
\Ytwoqubitsprobadirxxplusplusdiffminusminusfactorimag
\}$.
The only unknowns of
(\ref{eq-twoqubitsprobaplusplusdirxx-minus-twoqubitsprobaminusminusdirxx-def-and-explicit-mean})
are then
\ytwoqubitsprobadirxxplusplusdiffminusminusphasediffcosgen\
and
\ytwoqubitsprobadirxxplusplusdiffminusminusphasediffsingen.
One could try and solve a single equation
(\ref{eq-twoqubitsprobaplusplusdirxx-minus-twoqubitsprobaminusminusdirxx-def-and-explicit-mean}),
by taking into account that
\ytwoqubitsprobadirxxplusplusdiffminusminusphasediffcosgen\
and
\ytwoqubitsprobadirxxplusplusdiffminusminusphasediffsingen\
are the cosine and sine of the same angle
(see
(\ref{eq-def-twoqubitsprobadirxxplusplusdiffminusminusphasediffcosgen})-%
(\ref{eq-def-twoqubitsprobadirxxplusplusdiffminusminusphasediffsingen})).
However, the solutions of such an equation yield a
problematic indeterminacy.
This problem is avoided by creating \emph{two} linearly independent
equations
(\ref{eq-twoqubitsprobaplusplusdirxx-minus-twoqubitsprobaminusminusdirxx-def-and-explicit-mean}),
by using two sets of statistics for
\yparamqubitonestateplusmodulus , 
\yparamqubittwostateplusmodulus ,
$(
\Yparamqubitonestateminusphase
-
\Yparamqubitonestateplusphase )$
and
$
( \Yparamqubittwostateminusphase
-
\Yparamqubittwostateplusphase
)
$.
Solving these two equations yields
\ytwoqubitsprobadirxxplusplusdiffminusminusphasediffcosgen\
and
\ytwoqubitsprobadirxxplusplusdiffminusminusphasediffsingen.

\subsection{Estimating the quantum process}\label{sec-estimate-process}
We finally show how 
the estimates of
the parameters
\ytwoqubitresultphaseevolsin,
\ytwoqubitsprobadirxxplusplusdiffminusminusphasediffcosgen\
and
\ytwoqubitsprobadirxxplusplusdiffminusminusphasediffsingen\
obtained above
may be used
to estimate the matrix
\yopmixdiag\
of
(\ref{eq-opmixdiagdef})
and hence the complete matrix
\yopmix\
of 
(\ref{eq-opmixmatrixdecompose}),
which defines the considered process in the standard basis.

In a first method,
we only consider a single value of the time interval
$(t - t_0)$
involved in
(\ref{eq-opmixdiagdef}),
that we
hereafter
denote as
\ytwoqubitwritereadtimeintervalindexone.
Eq.
(\ref{eq-def-Ytwoqubitresultphaseevol-versiontwo})-%
(\ref{eq-def-Ytwoqubitresultphaseevol-versionthree})
may then be inverted as
\yeqdefexchangetensorppalvaluexyvstwoqubitresultphaseevolindexd
where
\ytwoqubitresultphaseevolindexd\ is one determination associated
with the actual value
\ytwoqubitresultphaseevol,
i.e. 
\ytwoqubitresultphaseevolindexd\
is equal to
\ytwoqubitresultphaseevol\
up to the additive constant
$- \Yexchangetensorppalvaluexyshiftindetermint
\pi$,
where
\yexchangetensorppalvaluexyshiftindetermint\
is an integer.
Similarly,
(\ref{eq-def-twoqubitsprobadirxxplusplusdiffminusminusphasediffcosgen})-%
(\ref{eq-def-twoqubitsprobadirxxplusplusdiffminusminusphasediffsingen})
and
(\ref{eq-twoqubitsprobadirxxplusplusdiffminusminusphasediff-vs-exchange-tensor})
may be inverted as
\yeqdefexchangetensorppalvaluezvstwoqubitsprobadirxxplusplusdiffminusminusphasediffindexd
where
\ytwoqubitsprobadirxxplusplusdiffminusminusphasediffindexd\
is one determination associated
with the actual value
\ytwoqubitsprobadirxxplusplusdiffminusminusphasediff,
i.e. 
\ytwoqubitsprobadirxxplusplusdiffminusminusphasediffindexd\
is equal to
\ytwoqubitsprobadirxxplusplusdiffminusminusphasediff\
up to the additive constant
$2
\Yexchangetensorppalvaluezshiftindetermint
\pi$,
where
\yexchangetensorppalvaluezshiftindetermint\
is an integer.

Eq.
(\ref{eq-exchangetensorppalvaluexy-vs-twoqubitresultphaseevolindexd})-%
(\ref{eq-def-twoqubitsprobadirxxplusplusdiffminusminusphasediffindexd})
define the expressions of the scaled \emph{actual}
principal
values
\yexchangetensorppalvaluexy\
and
\yexchangetensorppalvaluez\
with
respect to 
the
\emph{actual}
values of
\ytwoqubitresultphaseevolsin,
\ytwoqubitsprobadirxxplusplusdiffminusminusphasediffcosgen\
and
\ytwoqubitsprobadirxxplusplusdiffminusminusphasediffsingen.
The latter values are unknown but,
in practice, the procedure defined in Section
\ref{sec-estim-v}
yields an estimate
\ytwoqubitresultphaseevolsinestimtwo\
of the value of
\ytwoqubitresultphaseevolsin\
(for the considered value
\ytwoqubitwritereadtimeintervalindexone).
From this, one may derive an estimate
\ytwoqubitresultphaseevolindexdestim\
of
\ytwoqubitresultphaseevolindexd\
by using
\ytwoqubitresultphaseevolsinestimtwo\
in
(\ref{eq-twoqubitresultphaseevolindexd}).
One would then like to derive an 
estimate 
\yexchangetensorppalvaluexyestim\
of
\yexchangetensorppalvaluexy\
from 
(\ref{eq-exchangetensorppalvaluexy-vs-twoqubitresultphaseevolindexd}).
But one does not know the actual value
\yexchangetensorppalvaluexyshiftindetermint\
involved in
(\ref{eq-exchangetensorppalvaluexy-vs-twoqubitresultphaseevolindexd})
in the fully blind case considered here,
i.e. when no prior information is available about the
value of
\yexchangetensorppalvaluexy\
(as opposed to the case when one at least
knows in which range of values
\yexchangetensorppalvaluexy\
is situated, which defines the minimum and maximum possible
values of
\yexchangetensorppalvaluexyshiftindetermint).
In this blind method,
one can then only select an arbitrary
integer
\yexchangetensorppalvaluexyshiftindetermintestim\
and derive the corresponding
scaled ``shifted estimate''
of 
\yexchangetensorppalvaluexy\
by using
\yeqdefexchangetensorppalvaluexyvstwoqubitresultphaseevolindexdestim
Similarly, the procedure defined in Section
\ref{sec-estim-w-both}
yields estimates
\ytwoqubitsprobadirxxplusplusdiffminusminusphasediffcosgenestim\
and
\ytwoqubitsprobadirxxplusplusdiffminusminusphasediffsingenestim\
of the values of
\ytwoqubitsprobadirxxplusplusdiffminusminusphasediffcosgen\
and
\ytwoqubitsprobadirxxplusplusdiffminusminusphasediffsingen\
(for the considered value
\ytwoqubitwritereadtimeintervalindexone).
From this, one first derives an estimate 
\ytwoqubitsprobadirxxplusplusdiffminusminusphasediffindexdestim\
of
\ytwoqubitsprobadirxxplusplusdiffminusminusphasediffindexd\
by using
\ytwoqubitsprobadirxxplusplusdiffminusminusphasediffcosgenestim\
and
\ytwoqubitsprobadirxxplusplusdiffminusminusphasediffsingenestim\
in
(\ref{eq-def-twoqubitsprobadirxxplusplusdiffminusminusphasediffindexd}).
Then, based on
(\ref{eq-exchangetensorppalvaluez-vs-twoqubitsprobadirxxplusplusdiffminusminusphasediffindexd}),
one derives a scaled shifted estimate
\yexchangetensorppalvaluezestim\
of
\yexchangetensorppalvaluez\
by using
\yeqdefexchangetensorppalvaluezvstwoqubitsprobadirxxplusplusdiffminusminusphasediffindexdestim
where
\yexchangetensorppalvaluezshiftindetermintestim\
is an
arbitrarily selected
integer.
When neglecting estimation errors for
\ytwoqubitresultphaseevolsin,
\ytwoqubitsprobadirxxplusplusdiffminusminusphasediffcosgen\
and
\ytwoqubitsprobadirxxplusplusdiffminusminusphasediffsingen,
and hence for
\ytwoqubitresultphaseevolindexd\
and
\ytwoqubitsprobadirxxplusplusdiffminusminusphasediffindexd,
and when
taking the difference between
(\ref{eq-exchangetensorppalvaluexy-vs-twoqubitresultphaseevolindexd})
and
(\ref{eq-exchangetensorppalvaluexy-vs-twoqubitresultphaseevolindexd-estim}),
then
between
(\ref{eq-exchangetensorppalvaluez-vs-twoqubitsprobadirxxplusplusdiffminusminusphasediffindexd})
and
(\ref{eq-exchangetensorppalvaluez-vs-twoqubitsprobadirxxplusplusdiffminusminusphasediffindexd-estim}),
one
gets
\yeqdefexchangetensorppalvaluebothestimvsactual
The shifted estimates
$\frac{ 
\Yexchangetensorppalvaluexyestim
\Ytwoqubitwritereadtimeintervalindexone
} 
{ \hbar }$
and
$\frac{ 
\Yexchangetensorppalvaluezestim
\Ytwoqubitwritereadtimeintervalindexone
} 
{ \hbar }$
provided by this method are therefore equal
to the quantities of interest,
that is
$\frac{ 
\Yexchangetensorppalvaluexy
\Ytwoqubitwritereadtimeintervalindexone
} { \hbar }$
and
$\frac{ 
\Yexchangetensorppalvaluez
\Ytwoqubitwritereadtimeintervalindexone
} { \hbar }$,
only up to 
(the above neglected estimation errors and)
additive constants which are integer
multiples of
$\pi$.
These constants are the ``undeterminacies'' 
of this method
in the classical BSS
sense, i.e. the undesired remaining differences between the above
estimated and actual quantities,
from the point of view of the
quantities
$\frac{ 
\Yexchangetensorppalvaluexy
\Ytwoqubitwritereadtimeintervalindexone
} { \hbar }$
and
$\frac{ 
\Yexchangetensorppalvaluez
\Ytwoqubitwritereadtimeintervalindexone
} { \hbar }$.
They then yield the following indeterminacies from the point
of view of the matrix
\yopmix\ of the
considered quantum process,
which is eventually to be estimated.
Using the above estimates
$\frac{ 
\Yexchangetensorppalvaluexyestim
\Ytwoqubitwritereadtimeintervalindexone
} 
{ \hbar }$
and
$\frac{ 
\Yexchangetensorppalvaluezestim
\Ytwoqubitwritereadtimeintervalindexone
} 
{ \hbar }$,
one derives
the associated estimate
of the matrix
\yopmix\
(i) by inserting these 
estimates, which may be
expressed as
(\ref{eq-exchangetensorppalvaluexy-estim-vs-actual})-%
(\ref{eq-exchangetensorppalvaluez-estim-vs-actual}),
into
(\ref{eq-opmixdiagdef})-%
(\ref{eq-omega-zero-zero-express}),
which yields the corresponding 
estimate
\yopmixdiagestim\
of
\yopmixdiag ,
and (ii) finally by using
(\ref{eq-opmixmatrixdecompose})
and
(\ref{eq-opmixbasesdef})
to
derive the associated
estimate
of
\yopmix .
These calculations
especially
yield
\ymodifartitionehundredsixvonestepone{(taking into account that
$
e^{
\Ysqrtminusone
\Yexchangetensorppalvaluezshiftindetermintestimminusactual
2
\pi
}
=
1
$
and
$
e^{
\Yexchangetensorppalvaluexyshiftindetermintestimminusactual
2 \pi
}
=
1
$)}
\yeqdefopmixdiagestimvsopmixdiag

The estimate
\yopmixdiagestim\
provided by this first method is therefore equal to the actual matrix
\yopmixdiag\ up to the phase factor
$e^{
\Ysqrtminusone
(
\Yexchangetensorppalvaluezshiftindetermintestimminusactual
\pi
-
\Yexchangetensorppalvaluexyshiftindetermintestimminusactual
\frac{\pi}{2}
)
}$.
\ymodifartitionehundredsixvonestepone{More specifically, this factor
is equal to 1 and thus diseappears}
for part of the 
possible values of
the integers
\yexchangetensorppalvaluezshiftindetermintestimminusactual\
and
\yexchangetensorppalvaluexyshiftindetermintestimminusactual%
\ymodifartitionehundredsixvonestepone{, e.g. when
\yexchangetensorppalvaluezshiftindetermintestimminusactual\
is a multiple of 2
and
\yexchangetensorppalvaluexyshiftindetermintestimminusactual\
is a multiple of 4.}
This yields
the same phenomenon for
\yopmix .
\ymodifartitionehundredsixvonestepone{The general phase factor
$e^{
\Ysqrtminusone
(
\Yexchangetensorppalvaluezshiftindetermintestimminusactual
\pi
-
\Yexchangetensorppalvaluexyshiftindetermintestimminusactual
\frac{\pi}{2}
)
}$}
cannot be avoided with this method if no 
additional information is available.
It is 
\ymodifartitionehundredsixvonestepone{the only and
quite weak}
indeterminacy of this BQPT method from
the point of view of
\yopmixdiag\
and
\yopmix .
Moreover, we hereafter introduce an
extended version of that method, which completely removes
this indeterminacy by taking
the typical applications of (B)QPT methods
into account.

As discussed e.g. in
\cite{amq45},
\cite{amq50-physical-review},
\cite{booknielsen},
\cite{amq48},
\cite{amq52-physical-review},
\cite{paper-white-gilchrist-2007},
QPT (and hence our blind extension)
may especially be used as
a
tool for characterizing quantum gates,
which are 
the building blocks
of 
quantum computers.
This characterization 
is typically
performed before
using the considered
gates for 
quantum computation,
thus leading to
a two-phase approach, composed of an
``identification phase'' and then of a 
``computation phase'',
for these
quantum processes/gates.
Moreover, one may consider scenarios where these
processes/gates
are used
in 
coherent
but somewhat
different conditions during the
identification and computation phases.
We hereafter propose such an approach for extending the above
BQPT method so as to remove its indeterminacy.
We do not claim
that the Heisenberg coupling
model considered in this paper could be used as a suitable 
process/gate for
quantum computers:
it is just used as an example
hereafter, to illustrate a possible
procedure for removing BQPT indeterminacies, thus then allowing
the reader to extend this procedure to other processes/gates
that could be of interest in other configurations.

The approach that we propose uses three
values of the time interval
$(t - t_0)$
involved in
(\ref{eq-opmixdiagdef}),
that we
hereafter
denote as
\ytwoqubitwritereadtimeintervalindexone,
\ytwoqubitwritereadtimeintervalindextwo\
and
\ytwoqubitwritereadtimeintervalindexthree.
The first step of
the identification phase
uses the time interval
\ytwoqubitwritereadtimeintervalindexone%
\ymodifartitionehundredsixvonestepone{, essentially to obtain
an estimate of
\yexchangetensorppalvaluexy\ associated with this value
\ytwoqubitwritereadtimeintervalindexone ,
that we hereafter denote as
\yexchangetensorppalvaluexyestimargtwoqubitwritereadtimeintervalindexone\
for the sake of clarity.
More precisely, this
first step of
the identification phase}
derives the shifted estimate
$\frac{ 
\ymodifartitionehundredsixvonestepone{\Yexchangetensorppalvaluexyestimargtwoqubitwritereadtimeintervalindexone}
\Ytwoqubitwritereadtimeintervalindexone
} 
{ \hbar }$
in the same way as in the above first BQPT method,
i.e. using
(\ref{eq-exchangetensorppalvaluexy-vs-twoqubitresultphaseevolindexd-estim})%
\ymodifartitionehundredsixvonestepone{, with
\yexchangetensorppalvaluexyestim\
here replaced by
\yexchangetensorppalvaluexyestimargtwoqubitwritereadtimeintervalindexone.
Therefore,
when neglecting estimation errors,
this
again yields
(\ref{eq-exchangetensorppalvaluexy-estim-vs-actual}),
but with our modified notations, that is
\yeqdefexchangetensorppalvaluebothestimvsactualargtwoqubitwritereadtimeintervalindexone}
The second step of
the identification phase
\ymodifartitionehundredsixvonestepone{then
uses} the time interval
\ytwoqubitwritereadtimeintervalindextwo,
with
$
\Ytwoqubitwritereadtimeintervalindextwo
=
2
\Ytwoqubitwritereadtimeintervalindexone
$
(\ytwoqubitwritereadtimeintervalindextwo\
may instead be set to
any other even multiple of
\ytwoqubitwritereadtimeintervalindexone,
but we keep the values of
\ytwoqubitwritereadtimeintervalindexone,
\ytwoqubitwritereadtimeintervalindextwo\
and
\ytwoqubitwritereadtimeintervalindexthree\
as close as possible to one another, in
order to minimize the differences in the
conditions of operation in the two steps
of the identification phase and in the
computation phase).
\ymodifartitionehundredsixvonestepone{This second step of
the identification phase
essentially aims at obtaining
an estimate of
\yexchangetensorppalvaluez\ associated with the value
\ytwoqubitwritereadtimeintervalindextwo ,
that we therefore hereafter denote as
\yexchangetensorppalvaluezestimargtwoqubitwritereadtimeintervalindextwo .
More precisely, this
second step}
derives the shifted estimate
$\frac{ 
\ymodifartitionehundredsixvonestepone{\Yexchangetensorppalvaluezestimargtwoqubitwritereadtimeintervalindextwo}
\Ytwoqubitwritereadtimeintervalindextwo
} 
{ \hbar }$
in the same way as 
$\frac{ 
\Yexchangetensorppalvaluezestim
\Ytwoqubitwritereadtimeintervalindexone
} 
{ \hbar }$
in the above first BQPT method,
except that
\ymodifartitionehundredsixvonestepone{this step is here performed with
\ytwoqubitwritereadtimeintervalindextwo,
so that it}
uses
(\ref{eq-exchangetensorppalvaluez-vs-twoqubitsprobadirxxplusplusdiffminusminusphasediffindexd-estim})
with
\ytwoqubitwritereadtimeintervalindexone\
replaced by
\ytwoqubitwritereadtimeintervalindextwo,
\ymodifartitionehundredsixvonestepone{moreover}
taking into account that 
\ymodifartitionehundredsixvonestepone{the}
term
$\frac{ 
\Yexchangetensorppalvaluexyestim
\Ytwoqubitwritereadtimeintervalindextwo
} 
{ \hbar }$
\ymodifartitionehundredsixvonestepone{of
this modified version of
(\ref{eq-exchangetensorppalvaluez-vs-twoqubitsprobadirxxplusplusdiffminusminusphasediffindexd-estim})}
is here obtained as being equal to 
\ymodifartitionehundredsixvonestepone{the}
value
$\frac{ 
\ymodifartitionehundredsixvonestepone{\Yexchangetensorppalvaluexyestimargtwoqubitwritereadtimeintervalindexone}
\Ytwoqubitwritereadtimeintervalindexone
} 
{ \hbar }$
\ymodifartitionehundredsixvonestepone{of this second BQPT method}
multiplied by 2.
When neglecting estimation errors,
taking the difference between
the modified versions of
(\ref{eq-exchangetensorppalvaluez-vs-twoqubitsprobadirxxplusplusdiffminusminusphasediffindexd})
and
(\ref{eq-exchangetensorppalvaluez-vs-twoqubitsprobadirxxplusplusdiffminusminusphasediffindexd-estim}),
and using
\ymodifartitionehundredsixvonestepone{(\ref{eq-exchangetensorppalvaluexy-estim-vs-actual-argtwoqubitwritereadtimeintervalindexone})},
Eq.
(\ref{eq-exchangetensorppalvaluez-estim-vs-actual})
is thus replaced by
\yeqdefexchangetensorppalvaluezestimvsactualtwoqubitwritereadtimeintervalindextwo
The computation phase
then involves the same type of quantum process
(\ref{eq-opmixmatrixdecompose})-%
(\ref{eq-opmixdiagdef}),
but with a time interval
$
(
t - t_0
)
$,
between
input state preparation at time
$
t_0
$
and
output state use at
time
$
t
$,
which is set to
\ytwoqubitwritereadtimeintervalindexthree,
with
$
\Ytwoqubitwritereadtimeintervalindexthree
=
2
\Ytwoqubitwritereadtimeintervalindextwo
$
(again, \ytwoqubitwritereadtimeintervalindexthree\
may instead be set to
any other even multiple of
\ytwoqubitwritereadtimeintervalindextwo,
but we keep the values of
\ytwoqubitwritereadtimeintervalindexone,
\ytwoqubitwritereadtimeintervalindextwo\
and
\ytwoqubitwritereadtimeintervalindexthree\
as close as possible to one another).
This computation phase should then be analyzed as follows.
During that phase,
the considered 
actual process is defined by
(\ref{eq-opmixmatrixdecompose})-%
(\ref{eq-omega-zero-zero-express}),
but with
$(t - t_0)$
replaced by
\ytwoqubitwritereadtimeintervalindexthree.
From the point of view of that computation phase,
the estimate of that actual process
is obtained by replacing
$\frac{ 
\Yexchangetensorppalvaluexy
(t - t_0)
} { \hbar }$
and
$\frac{ 
\Yexchangetensorppalvaluez
(t - t_0)
} { \hbar }$
by
$\frac{ 
\ymodifartitionehundredsixvonestepone{\Yexchangetensorppalvaluexyestimargtwoqubitwritereadtimeintervalindexone}
\Ytwoqubitwritereadtimeintervalindexthree
} 
{ \hbar }$
and
$\frac{ 
\ymodifartitionehundredsixvonestepone{\Yexchangetensorppalvaluezestimargtwoqubitwritereadtimeintervalindextwo}
\Ytwoqubitwritereadtimeintervalindexthree
} 
{ \hbar }$
in
(\ref{eq-opmixmatrixdecompose})-%
(\ref{eq-omega-zero-zero-express}),
the latter estimates being
derived
as explained above by our extended BQPT method
(up to the factors
$\Ytwoqubitwritereadtimeintervalindexthree /
\Ytwoqubitwritereadtimeintervalindexone$
and
$\Ytwoqubitwritereadtimeintervalindexthree /
\Ytwoqubitwritereadtimeintervalindextwo$).
These estimates have the properties defined by
\ymodifartitionehundredsixvonestepone{(\ref{eq-exchangetensorppalvaluexy-estim-vs-actual-argtwoqubitwritereadtimeintervalindexone})}
and
(\ref{eq-exchangetensorppalvaluez-estim-vs-actual-twoqubitwritereadtimeintervalindextwo}).
Since
$\Ytwoqubitwritereadtimeintervalindexthree
/
\Ytwoqubitwritereadtimeintervalindexone
=
4$
and
$\Ytwoqubitwritereadtimeintervalindexthree
/
\Ytwoqubitwritereadtimeintervalindextwo
=
2$,
this yields
\yeqdefexchangetensorppalvaluezestimvsactualtwoqubitwritereadtimeintervalindexthree
Comparing these expressions with
(\ref{eq-exchangetensorppalvaluexy-estim-vs-actual})
and
(\ref{eq-exchangetensorppalvaluez-estim-vs-actual})
shows that this second BQPT method it equivalent to the first one presented
in this section, except that, from the point
of view of the computation phase, (i) it use the
time interval
\ytwoqubitwritereadtimeintervalindexthree\
and
(ii)
\yexchangetensorppalvaluexyshiftindetermintestimminusactual\
and
\yexchangetensorppalvaluezshiftindetermintestimminusactual\
are respectively replaced by
$4
\Yexchangetensorppalvaluexyshiftindetermintestimminusactual$
and
$2
\Yexchangetensorppalvaluezshiftindetermintestimminusactual
$.
The analysis provided above for the first method therefore
also applies here when taking the above modifications into
account. In particular,
(\ref{yeq-opmixdiagestim-vs-opmixdiag}) also applies here,
but its phase factor 
$e^{
\Ysqrtminusone
(
\Yexchangetensorppalvaluezshiftindetermintestimminusactual
\pi
-
\Yexchangetensorppalvaluexyshiftindetermintestimminusactual
\frac{\pi}{2}
)
}$
here becomes
$e^{
\Ysqrtminusone
(
2
\Yexchangetensorppalvaluezshiftindetermintestimminusactual
\pi
-
2
\Yexchangetensorppalvaluexyshiftindetermintestimminusactual
\ymodifartitionehundredsixvonestepone{\pi}
)
}$
and is therefore always equal to one.
In other words, this extended BQPT method is
equivalent to forcing
\yexchangetensorppalvaluexyshiftindetermintestimminusactual\
and
\yexchangetensorppalvaluezshiftindetermintestimminusactual\
to be respectively equal to multiples of 4 and 2
from the point of view of the computation phase,
which suppresses the 
indeterminacy that the first method has in that phase.

The above discussion first means
that any value of
\yexchangetensorppalvaluexyshiftindetermintestimminusactual\
and hence
\yexchangetensorppalvaluexyshiftindetermintestim\
may be used during the identification phase of
our extended BQPT method.
From a practical point of view, 
the simplest implementation of this method therefore consists of selecting
$
\Yexchangetensorppalvaluexyshiftindetermintestim
=
0
$
in
\ymodifartitionehundredsixvonestepone{the modified version of}
(\ref{eq-exchangetensorppalvaluexy-vs-twoqubitresultphaseevolindexd-estim}),
i.e. it consists of setting
the estimate
$
\frac{ 
\ymodifartitionehundredsixvonestepone{\Yexchangetensorppalvaluexyestimargtwoqubitwritereadtimeintervalindexone}
\Ytwoqubitwritereadtimeintervalindexone
} 
{ \hbar }
$
to
$
-
\Ytwoqubitresultphaseevolindexdestim
$.
This estimate is then multiplied by
$\Ytwoqubitwritereadtimeintervalindexthree
/
\Ytwoqubitwritereadtimeintervalindexone
=
4$
when considering it from the point of view of the computation phase.
Similarly, during the identification phase
one
sets
the estimate
$\frac{ 
\ymodifartitionehundredsixvonestepone{\Yexchangetensorppalvaluezestimargtwoqubitwritereadtimeintervalindextwo}
\Ytwoqubitwritereadtimeintervalindextwo
} 
{ \hbar }$
by
using
(\ref{eq-exchangetensorppalvaluez-vs-twoqubitsprobadirxxplusplusdiffminusminusphasediffindexd-estim}),
with
\ytwoqubitwritereadtimeintervalindexone\
replaced by
\ytwoqubitwritereadtimeintervalindextwo\
and with
\yexchangetensorppalvaluezshiftindetermintestim\
preferably set to 0
(and with
$
\frac{ 
\Yexchangetensorppalvaluexyestim
\Ytwoqubitwritereadtimeintervalindextwo
} 
{ \hbar }
$
obtained by multiplying the above estimate
$
\frac{ 
\ymodifartitionehundredsixvonestepone{\Yexchangetensorppalvaluexyestimargtwoqubitwritereadtimeintervalindexone}
\Ytwoqubitwritereadtimeintervalindexone
} 
{ \hbar }
$
by
$\Ytwoqubitwritereadtimeintervalindextwo
/
\Ytwoqubitwritereadtimeintervalindexone
=
2$).
This estimate 
$\frac{ 
\ymodifartitionehundredsixvonestepone{\Yexchangetensorppalvaluezestimargtwoqubitwritereadtimeintervalindextwo}
\Ytwoqubitwritereadtimeintervalindextwo
} 
{ \hbar }$
is then multiplied by
$\Ytwoqubitwritereadtimeintervalindexthree
/
\Ytwoqubitwritereadtimeintervalindextwo
=
2$
when considering it from the point of view of the computation phase.

\section{Test results}\label{sec-tests}
The
physical implementation of qubits is an emerging
topic which is 
beyond the scope of this 
paper.
We therefore
assessed
the performance of the extended 
BQPT method proposed above
by means of numerical tests performed with
data derived from
a software simulation of the 
considered configuration.
Each elementary
test
consists of 
the following
stages. 
We first
create
a set of
\ytwoqubitseqnb\
input states
\ymixsyststateinitial.
Each such state is obtained
by randomly drawing its 
six parameters
\yparamqubitindexstdstateplusmodulus,
\yparamqubitindexstdstateplusphase\
and
\yparamqubitindexstdstateminusphase,
with
$\Yqubitindexstd \in \{ 1 , 2 \}$,
and then using
(\ref{eq-def-qubit-polar-qubit-indexstd}),
(\ref{eq-paramqubitindexstdstateminusmodulus-vs-paramqubitindexstdstateplusmodulus}),
(\ref{eq-etat-deuxspin-plusplus-initial-decompos})
\ymodifartitionehundredsixvonestepone{(the state
(\ref{eq-etat-deuxspin-plusplus-initial-decompos})
is defined by the above six parameters, but only
the four parameters 
\yparamqubitindexstdstateplusmodulus\
and
$
\Yparamqubitindexstdstateminusphase
-
\Yparamqubitindexstdstateplusphase
$
have a physical meaning).}
We then
process
the
states
\ymixsyststateinitial\
according to
(\ref{eq-statetfinalvsopmixstatetinit-components}),
with given values of the parameters of
the matrix
\yopmix\ which defines the quantum process to be identified.
This
yields the states 
\ymixsyststatefinal.
More precisely, we eventually
use simulated measurements of spin components
associated with these states 
\ymixsyststatefinal.
For measurements along the $Oz$ axis, this means that
we use the model
(\ref{eq-statecoef-vs-twoqubitsprobaplusplus-polar})-(\ref{eq-statecoef-vs-twoqubitsprobaminusminus-polar-vs-paramqubitindexstdstateplusmodulus})
with a given 
value of the
mixing parameter
\ytwoqubitresultphaseevolsin,
corresponding to the above values of the parameters of the matrix
\yopmix.
For each of the
\ytwoqubitseqnb\
states
\ymixsyststateinitial,
corresponding to parameter values
$(
\Yparamqubitonestateplusmodulus ,
\Yparamqubittwostateplusmodulus ,
\Ytwoqubitresultphaseinit
)$,
Eq.
(\ref{eq-statecoef-vs-twoqubitsprobaplusplus-polar})-%
(\ref{eq-statecoef-vs-twoqubitsprobaminusminus-polar-vs-paramqubitindexstdstateplusmodulus})
thus yield the corresponding set of probability values
$(
\Ytwoqubitsprobaplusplusdirzz
,
\Ytwoqubitsprobaplusminusdirzz
,
\Ytwoqubitsprobaminusminusdirzz
)$,
which are used as follows.
We use \ywritereadonestatenb\
prepared copies of
the considered state
\ymixsyststateinitial\
to
simulate \ywritereadonestatenb\
random-valued
two-qubit
spin component measurements along the
$Oz$ axis,
drawn with the above probabilities
$(
\Ytwoqubitsprobaplusplusdirzz
,
\Ytwoqubitsprobaplusminusdirzz
,
\Ytwoqubitsprobaminusminusdirzz
)$.
We then derive the 
sample 
frequencies 
of the results of 
these
\ywritereadonestatenb\
measurements,
which are
estimates
of
\ytwoqubitsprobaplusplusdirzz,
\ytwoqubitsprobaplusminusdirzz\
and
\ytwoqubitsprobaminusminusdirzz\
for the considered 
state
\ymixsyststateinitial\
(see
(\ref{eq-def-qubitnbarbspaceoveralleventindexequalstdwithvecbasisprobseqindexapproxwritereadonestatenb})).
Then
computing the averages 
of these 
\ywritereadonestatenb-preparation
estimates over all
\ytwoqubitseqnb\
source vectors 
\ymixsyststateinitial\
yields
$(
\Ytwoqubitseqnb
\Ywritereadonestatenb
)$-preparation
estimates
of probability expectations
$E \{ 
\Ytwoqubitsprobaindexstddirzz \}$
(see  (\ref{eq-def-qubitnbarbspaceoveralleventindexequalstdwithvecbasisprobexpectapproxtwotwoqubitseqnb})).
Spin component measurements
for the $Ox$ axis are handled similarly,
thus yielding 
estimates
of probability expectations
$E \{ 
\Ytwoqubitsprobaindexstddirxx \}$.
Both types of estimates
of probability expectations
are then used by our extended BQPT method,
as explained in the previous sections,
to derive
the estimates
$
\frac{ 
\Yexchangetensorppalvaluexyestim
\Ytwoqubitwritereadtimeintervalindexone
} 
{ \hbar }
$
and
$\frac{ 
\Yexchangetensorppalvaluezestim
\Ytwoqubitwritereadtimeintervalindextwo
} 
{ \hbar }$,
from which we then derive the estimates of
\yopmixdiag\
and eventually
\yopmix\
corresponding to the computation phase
that uses the time interval
\ytwoqubitwritereadtimeintervalindexthree.

In these tests,
the above
parameters 
\ytwoqubitseqnb\
and
\ywritereadonestatenb\
were varied as described further in this section,
whereas
the numerical values of the other parameters
were fixed
as explained in Appendix \ref{sec-appendix-test-conditions},
so that we used
the same values for the parameters
\ytwoqubitresultphaseevolsin,
\ytwoqubitsprobadirxxplusplusdiffminusminusphasediffcosgen\
and
\ytwoqubitsprobadirxxplusplusdiffminusminusphasediffsingen\
and for the
matrix
\yopmix\ in all tests.
For each considered set of conditions defined by the values of
\ytwoqubitseqnb\
and
\ywritereadonestatenb,
we performed
100 above-defined elementary
tests, with different sets of states
\ymixsyststateinitial, in order to assess
the statistical performance of the considered BQPT method
over 100 estimations of the same set of parameter values.
The performance criteria used to this end are defined as follows.
Separately for each of the scalar parameters
\ytwoqubitresultphaseevolsin,
\ytwoqubitsprobadirxxplusplusdiffminusminusphasediffcosgen\
and
\ytwoqubitsprobadirxxplusplusdiffminusminusphasediffsingen,
we computed the Normalized Root Mean Square Error (NRMSE) of that
parameter over all 100 estimations, defined as the ratio of its
RMSE to its actual value.
For the \emph{matrix} \yopmix, we first derived a \emph{scalar}
relative
error for each test, defined as the ratio of the Frobenius norm
of
the ``error matrix''
$
(
\Yopmixestim
-
\Yopmix
)
$,
where
\yopmixestim\
is the estimate of
\yopmix\
provided by our BQPT method,
to the Frobenius norm of
the actual
matrix
\yopmix\
(the Frobenius norm of a matrix
$A$
with entries
$a_{ij}$
is defined as
$\displaystyle \sqrt{\sum_i \sum_j a_{ij}^2}$).
We then computed the average of this relative error over 
all 100 estimations.

\begin{figure}[ht]
\centerline{\epsfig{file=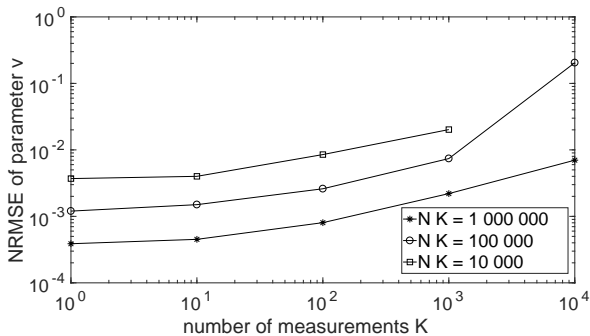,width=\linewidth}}
\caption{Normalized Root Mean Square
Error (NRMSE) of
estimation of parameter
\ytwoqubitresultphaseevolsin\ vs. number
\ywritereadonestatenb\
of preparations of each of the
\ytwoqubitseqnb\
used
states.}
\label{fig-NRMSE-v}
\end{figure}

\begin{figure}[ht]
\centerline{\epsfig{file=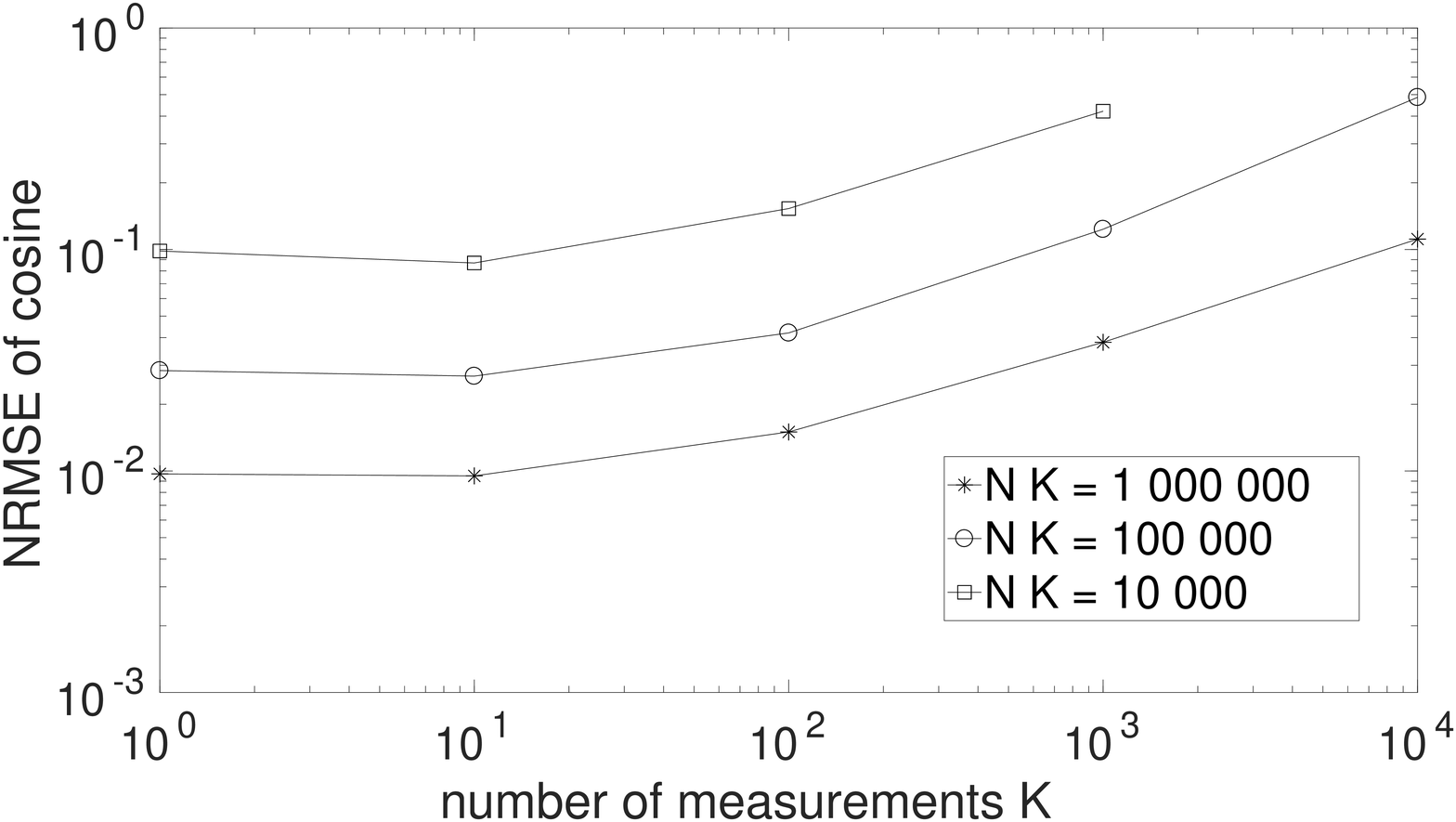,width=\linewidth}}
\caption{NRMSE of
estimation of parameter
\ytwoqubitsprobadirxxplusplusdiffminusminusphasediffcosgen\
vs. number
\ywritereadonestatenb\
of preparations of each of the
\ytwoqubitseqnb\
used
states.}
\label{fig-NRMSE-cos}
\end{figure}

\begin{figure}[ht]
\centerline{\epsfig{file=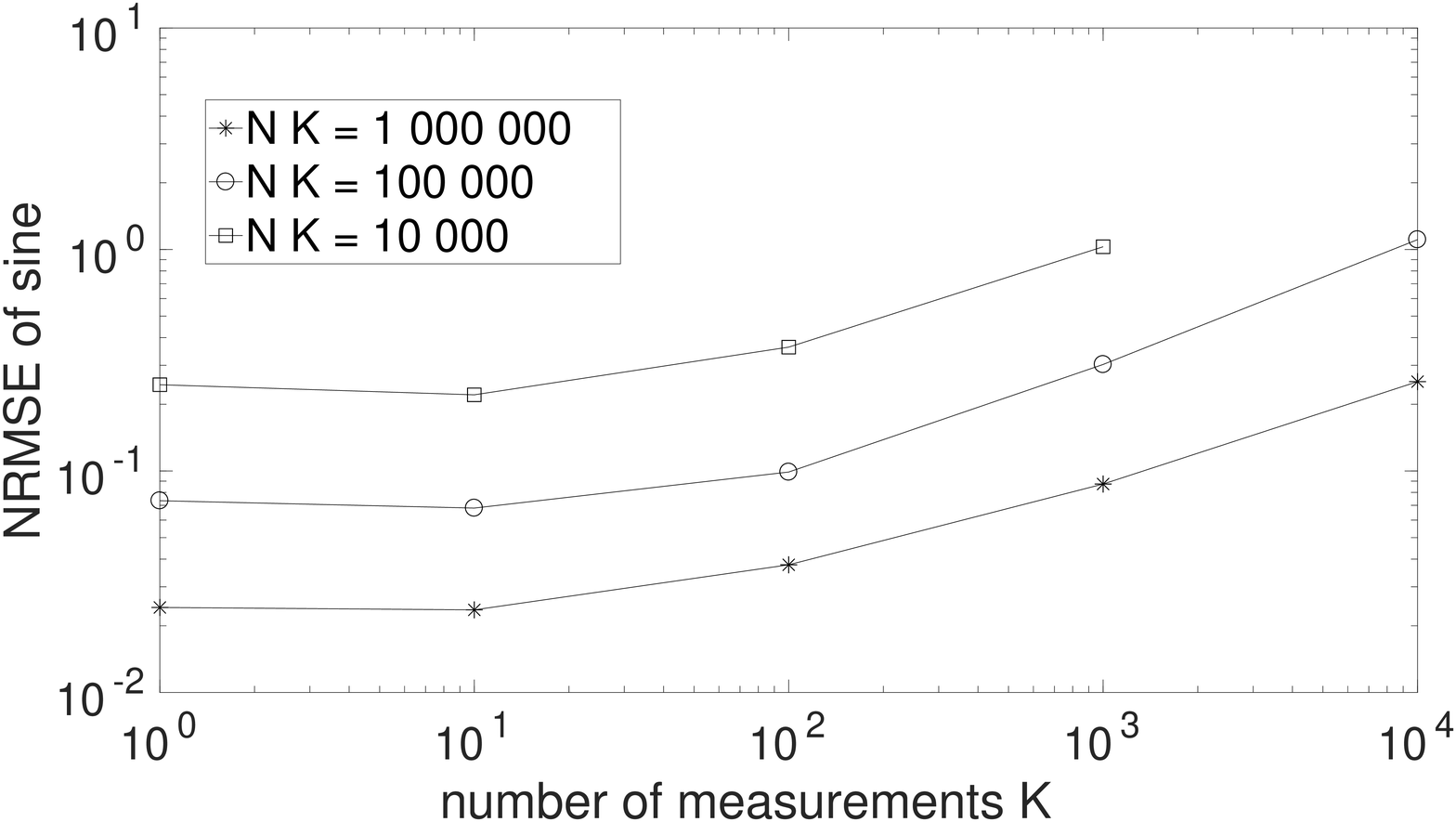,width=\linewidth}}
\caption{NRMSE of
estimation of parameter
\ytwoqubitsprobadirxxplusplusdiffminusminusphasediffsingen\
vs. number
\ywritereadonestatenb\
of preparations of each of the
\ytwoqubitseqnb\
used
states.}
\label{fig-NRMSE-sin}
\end{figure}

\begin{figure}[ht]
\centerline{\epsfig{file=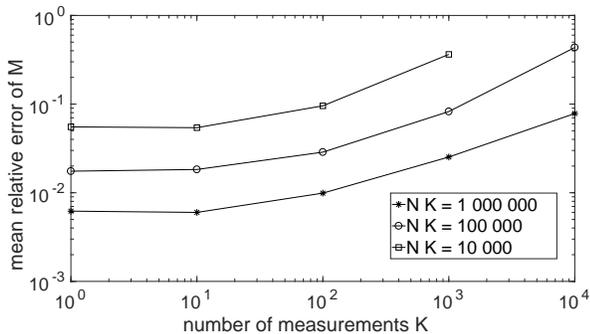,width=\linewidth}}
\caption{Mean relative error of
estimation of matrix \yopmix\
vs. number
\ywritereadonestatenb\
of preparations of each of the
\ytwoqubitseqnb\
used
states.}
\label{fig-NRMSE-M}
\end{figure}

The values of these four performance criteria are shown in
Fig. \ref{fig-NRMSE-v}
to \ref{fig-NRMSE-M},
where
each plot corresponds to a fixed value of the product
$
\Ytwoqubitseqnb
\Ywritereadonestatenb
$,
i.e. of the complexity
of the BQPT method in terms of the total number of state preparations.
Each plot shows the variations of the 
considered performance criterion
vs. 
\ywritereadonestatenb,
hence with 
\ytwoqubitseqnb\
varied accordingly,
to keep the considered fixed value of
$
\Ytwoqubitseqnb
\Ywritereadonestatenb
$.
This first shows 
that the proposed BQPT method is able to operate
with a number
\ywritereadonestatenb\
of preparations 
per state
\ymixsyststateinitial\
decreased down to one, as expected.
Moreover, 
for a fixed value of
$
\Ytwoqubitseqnb
\Ywritereadonestatenb
$,
the errors
improve when
\ywritereadonestatenb\
decreases, 
which is expected to be due to the fact that the
number
\ytwoqubitseqnb\
of \emph{different} used states thus increases,
allowing the estimation method to better
explore the statistics of the considered random process.
Thus 
using
$
\Ywritereadonestatenb
=
1
$,
the mean relative error for the matrix
\yopmix\
defining the considered quantum process
\ymodifartitionehundredsixvonestepone{(see Fig. \ref{fig-NRMSE-M})}
can e.g. here be made equal to 
5.53 \%
for
$\Ytwoqubitseqnb
=
10^4$
or
1.75 \%
for
$\Ytwoqubitseqnb
=
10^5$
or
0.62 \%
for
$\Ytwoqubitseqnb
=
10^6$.
In these tests, we used a simple protocol,
i.e. we considered the same values of
\ywritereadonestatenb\
and
\ytwoqubitseqnb\
in the six series of state preparations used for estimating all
parameters
(\ytwoqubitresultphaseevolsin,
\ytwoqubitsprobadirxxplusplusdiffminusminusphasediffcosgen\
and
\ytwoqubitsprobadirxxplusplusdiffminusminusphasediffsingen),
so that the total number of preparations is equal to
$6 
\Ytwoqubitseqnb
\Ywritereadonestatenb
$.
Different values of
\ywritereadonestatenb\
and
\ytwoqubitseqnb\
might be used
in these six series of state preparations, in order to optimize
the total number of preparations used to achieve a given error
for
\yopmix .
In particular,
when estimating the \emph{sign} of
\ytwoqubitresultphaseevolsin , the result is a binary decision,
not a continuous value which should be
accurately estimated, so that this sign could be obtained without
errors with a significantly lower number of state preparations,
thus making the total number of state preparations closer to
$5 
\Ytwoqubitseqnb
\Ywritereadonestatenb
$.
Besides,
Fig.
\ref{fig-NRMSE-v}
to
\ref{fig-NRMSE-sin}
show that, when using the same values of
\ywritereadonestatenb\
and
\ytwoqubitseqnb,
the parameter
\ytwoqubitresultphaseevolsin\
is estimated much more accurately than
\ytwoqubitsprobadirxxplusplusdiffminusminusphasediffcosgen\
and
\ytwoqubitsprobadirxxplusplusdiffminusminusphasediffsingen .
This is reasonable, because the measurements along the $Oz$ axis,
which are used to estimate
\ytwoqubitresultphaseevolsin,
yield a simpler model and hence a simpler estimation procedure
than
the measurements along the $Ox$ (and $Oz$) axis,
which are used to estimate
\ytwoqubitsprobadirxxplusplusdiffminusminusphasediffcosgen\
and
\ytwoqubitsprobadirxxplusplusdiffminusminusphasediffsingen .
When aiming at optimizing the use of state preparations,
one may therefore think of reducing the number of state preparations
for estimating
\ytwoqubitresultphaseevolsin\
as compared with those used for estimating
\ytwoqubitsprobadirxxplusplusdiffminusminusphasediffcosgen\
and
\ytwoqubitsprobadirxxplusplusdiffminusminusphasediffsingen,
in order to balance the estimation accuracies for these parameters.
However, it is not guaranteed that the estimation accuracy for
\yopmix\ will thus be significantly improved: in
Fig. \ref{fig-NRMSE-v}
to \ref{fig-NRMSE-M},
the estimation accuracy for
\yopmix\
has in intermediate value between the accuracies achieved for the
parameters
\ytwoqubitresultphaseevolsin,
\ytwoqubitsprobadirxxplusplusdiffminusminusphasediffcosgen\
and
\ytwoqubitsprobadirxxplusplusdiffminusminusphasediffsingen\
upon which
\yopmix\
depends,
i.e.
the accuracy of
\yopmix\
is not limited by those of its ``worst parameters'',
namely
\ytwoqubitsprobadirxxplusplusdiffminusminusphasediffcosgen\
and
\ytwoqubitsprobadirxxplusplusdiffminusminusphasediffsingen,
but takes advantage of its best parameter
\ytwoqubitresultphaseevolsin.
Based on all above results and considerations, a typical performance
level to be eventually kept in mind for the matrix
\yopmix\
which defines the considered quantum process
is a
mean relative error
of around 1 \%
for around 500,000 state preparations.

\section{Discussion and conclusion}\label{sec-conclusion}
Non-blind or blind QPT
may be defined as the identification, i.e. estimation,
of a given quantum process or gate.
As discussed 
in Section \ref{sec-estimate-process},
this allows one to characterize
the \emph{actual} behavior of such a
gate,
so that
(B)QPT
is a major quantum information processing 
tool.
The
usual, i.e. non-blind,
version of
QPT requires one to know, 
hence to precisely control
(i.e. prepare), 
the 
specific
quantum states used as
inputs of the quantum gate to be characterized.
The blind version of this tool,
i.e. BQPT, 
which is the first contribution proposed in this paper,
then provides an
attractive
extension of 
QPT, since it allows one to use 
input 
quantum states whose
values are unknown and arbitrary, except that they are requested to
meet some general known properties.

Such blind approaches especially have two potential
applications.
The most natural one is when the input states of the considered
process indeed cannot be known.
Such methods could then be of interest for characterizing quantum
gates while they are operating
and when only their results (output states) are available
to the user who is to characterize them
(provided some output states
\ymixsyststatefinal\
are available to perform BQPT,
with adequate values
of the above-defined preparation-to-measurement 
time interval
$(t - t_0)$).
This on-line characterization may be useful
e.g. if the transform performed by a
quantum gate slowly evolves over time (e.g. due to aging) and must be
monitored,
by characterizing it from time to time.
Besides, 
BQPT may be
of even higher interest
in more standard configurations, 
\ymodifartitionehundredsixvonestepone{when} 
the process input states
may be prepared and 
known:
BQPT then avoids the 
complexity of
\emph{accurately preparing the specific
states} 
which are required by usual QPT methods, because
BQPT can use \emph{any}
input states
(which have the requested general properties).

The second constraint of usual QPT methods is that they require one
to be able to prepare many copies of the \emph{same} 
(known) input state, which is cumbersome.
As a second contribution in this paper, we proposed methods which
avoid this constraint, since they allow one to use one instance or
several copies of each considered input state (and they
provide even better performance when decreasing the number of preparations
per state to one and increasing the number of different states
accordingly, for a fixed total number of preparations, i.e. fixed
complexity).
Briefly, this quite attractive property is obtained because the
proposed methods
do not use the estimated probabilities separately associated which
each considered random state, but only the expectations of
these random probabilities.

It should also be noted that the solution provided by each of the
proposed BQPT methods is defined by a unique
set of closed-form expressions.
This avoids the issues of estimation methods that are based on
the numerical optimization of cost functions and that yield false
solutions when they get trapped into local minima of these cost
functions.
Morever, the proposed BQPT methods only require a limited
number of types of measurements (only spin component
measurements along the $Ox$ and $Oz$ axes in the case considered
here), which simplifies their practical use. This results from
the fact that these methods
only perform the types of measurements which are needed to
get enough information about the required unknown parameter values.
This should be contrasted with methods which use a larger
set of types of measurements to first completely
restore quantum states.

The main global result of this paper is therefore
the joint concept of Single-preparation Blind QPT (SBQPT).
It was here
illustrated and numerically validated for a specific type of
quantum process, based on cylindrical-symmetry Heisenberg coupling,
for the sake of clarity. However, it may be applied to a much
wider class of processes and to methods based on other statistical
parameters of quantum measurement outcomes or other quantum
state properties.
Such extensions will
be reported in future papers.

%% file: artiti106v1_appendix.tex
\section{From 
multiple-preparation
to
single-preparation
quantum information processing
(QIP)}
\label{sec-stochastic-QIP}
In this appendix, we first summarize the concepts and
notations that are used in conventional,
i.e. multiple-preparation, QIP
(see Section \ref{sec-stochastic-QIP-standard-concepts})
and that we need to then introduce our non-conventional,
single-preparation, approach to QIP
(see Section \ref{sec-stochastic-QIP-introduce})
for an arbitrary number of qubits.
\subsection{%
Multiple-preparation
\ymodifartitionehundredsixvonestepone{QIP}%
}
\label{sec-stochastic-QIP-standard-concepts}

\ymodifartitionehundredsixvonestepone{%
Throughout this paper,
qubits
are 
physically
implemented as 
spins
1/2. 
If such a
qubit%
,
with index
$
i
$%
,
is described with 
a
pure
{and 
{``deterministic''
quantum state%
, 
as defined in
\cite{amoi6-67}}%
}%
,
it is represented by a normalized vector of
a two-dimensional space \yqubitonespaceindexstd%
,
expressed
as
\begin{equation}
\label{eq-twoqubit-state-init-index-i-subscript-i}
\Yqubitonetimenonestateindexqubitoneindexstd
=
\alpha _i | + 
{\rangle}
_{\Yqubitoneindexstd}
+ \beta _i | - 
{\rangle}
_{\Yqubitoneindexstd}
\end{equation}
where
$ \alpha_i $ and $ \beta_i $
are two 
{fixed}
complex-valued
coefficients
constrained to 
meet the normalization condition for
\yqubitonetimenonestateindexqubitoneindexstd .%
}
The index
\yqubitoneindexstd\
in the above notations
$
| + 
\rangle
_{\Yqubitoneindexstd}
$
and
$
| -
\rangle
_{\Yqubitoneindexstd}
$
is most often
omitted in the literature, but we
keep it here,
to clarify the notations that we hereafter introduce for \emph{a set of}
qubits.

Let us now consider an arbitrary number
\yqubitnbarb\ of 
distinguishable
\cite{amoi6-64}
qubits, with indices
$
\Yqubitoneindexstd
\in
\{
1,
\dots
,
\Yqubitnbarb
\}
$.
If the state
\yqubitnbarbtimenonestate\ of this 
{set}
of qubits is pure
{and deterministic}%
, 
it belongs to
the 
space
\yqubitbothspace\
defined as the tensor product
(denoted
as
$
\otimes
$)
of
the 
above spaces
\ymodifartitionehundredsixvonestepone{\yqubitonespaceindexstd .}
{The standard}
basis of
\yqubitbothspace\
consists of the
\ymodifartitionehundredsixvonestepone{$
2
^
{
\Yqubitnbarb
}
$
vectors
\mbox{$
| + 
\rangle
_{1}
\otimes
| + 
\rangle
_{2}
\otimes
\dots
\otimes
| + 
\rangle
_{\Yqubitnbarb - 1}
\otimes
| + 
\rangle
_{\Yqubitnbarb}
$}
to
\mbox{$
| - 
\rangle
_{1}
\otimes
| - 
\rangle
_{2}
\otimes
\dots
\otimes
| - 
\rangle
_{\Yqubitnbarb - 1}
\otimes
| - 
\rangle
_{\Yqubitnbarb}
$}}
that we hereafter respectively denote as
\yqubitnbarbspaceoverallvecbasiscoefindexequalstd ,
with
$
\Yqubitnbarbstatecoefindexequalstd
\in
\{
1,
\dots
,
2
^
{
\Yqubitnbarb
}
\}
$.
The 
{state}
of 
this set of
qubits
then reads
\begin{equation}
\Yqubitnbarbtimenonestate
=
\sum
_{
\Yqubitnbarbstatecoefindexequalstd
=
1
}
^{
2^{\Yqubitnbarb}
}
\Yqubitnbarbspaceoverallcoefindexequalstdwithvecbasis
\Yqubitnbarbspaceoverallvecbasiscoefindexequalstd
\label{eq-def-qubitnbarbtimenonestate}
\end{equation}
where
the 
complex-valued coefficients
\yqubitnbarbspaceoverallcoefindexequalstdwithvecbasis\
{are again fixed and arbitrary, 
except that
they}
meet the normalization condition
\begin{equation}
\sum
_{
\Yqubitnbarbstatecoefindexequalstd
=
1
}
^{
2^{\Yqubitnbarb}
}
|
\Yqubitnbarbspaceoverallcoefindexequalstdwithvecbasis
|
^2
=
1
\label{eq-qubitnbarbspaceoverallcoefindexequalstdwithvecbasis-normalization-cond}
.
\end{equation}

\ymodifartitionehundredsixvonestepone{%
The result 
obtained for one measurement 
of the 
spin component 
$s_{zi}$ 
of 
$\overrightarrow{s_{i}}$ 
along the quantization axis, 
for a single qubit
$i$ 
which
is in state 
(\ref{eq-twoqubit-state-init-index-i-subscript-i})
(see e.g.
\cite{amoi6-18}
{for}
details),
has
a random nature
and is
$
+
\frac{1}{2}
$
or
$
-
\frac{1}{2}
$
in normalized units.
The probabilities of obtaining these%
}
two values are respectively equal to
$
|
\alpha _i 
|
^2
$
and
$
|
\beta _i
|
^2
$,
that is, to 
the squared moduli
of the coefficients in
(\ref{eq-twoqubit-state-init-index-i-subscript-i})
which correspond to the vectors
$
| + 
\rangle
_{\Yqubitoneindexstd}
$
and
$
| -
\rangle
_{\Yqubitoneindexstd}
$
that are respectively associated with the allowed values
$
+
\frac{1}{2}
$
and
$
-
\frac{1}{2}
$.

When simultaneously performing such 
a measurement for each of the
qubits \yqubitoneindexstd\ of an overall set of \yqubitnbarb\
qubits,
the obtained result is a vector of \yqubitnbarb\ values.
The 
$
2
^
{\Yqubitnbarb}
$
possible values of this vector
are
$
[
+
\frac{1}{2}
,
+
\frac{1}{2}
,
\dots
,
+
\frac{1}{2}
,
+
\frac{1}{2}
]
$,
$
[
+
\frac{1}{2}
,
+
\frac{1}{2}
,
\dots
,
+
\frac{1}{2}
,
-
\frac{1}{2}
]
$,
and so on,
these values being respectively associated with the 
\ymodifartitionehundredsixvonestepone{above-defined
$
2
^
{
\Yqubitnbarb
}
$
basis vectors
\yqubitnbarbspaceoverallvecbasiscoefindexequalstd}
and hereafter indexed by 
\ymodifartitionehundredsixvonestepone{\yqubitnbarbstatecoefindexequalstd .}
Thus, the experiment consisting of this \yqubitnbarb -qubit
measurement yields a random result, and 
each 
elementary
event
\cite{book-papoulis}
\yqubitnbarbspaceoveralleventindexequalstdwithvecbasis\
is defined as:
the result of the experiment is equal to the
\yqubitnbarbstatecoefindexequalstd -th
\yqubitnbarb -entry vector in the above series
of possible values 
$
[
+
\frac{1}{2}
,
+
\frac{1}{2}
,
\dots
,
+
\frac{1}{2}
,
+
\frac{1}{2}
]
$
and so on.
Moreover, the probabilities of these 
events
are defined according to the 
principle 
{presented}
above for one qubit, that is
\begin{equation}
\Yqubitnbarbspaceoveralleventindexequalstdwithvecbasisprob
=
|
\Yqubitnbarbspaceoverallcoefindexequalstdwithvecbasis
|
^2
\hspace{10mm}
\forall
\
\Yqubitnbarbstatecoefindexequalstd
\in
\{
1,
\dots
,
2
^
{
\Yqubitnbarb
}
\}
.
\label{eq-def-qubitnbarbspaceoveralleventindexequalstdwithvecbasisprob}
\end{equation}

The
simplest
procedure,
applied
in practice to estimate the above probabilities
for a given \yqubitnbarb -qubit state,
uses a large number
(typically from a few 
{thousand}
up to one hundred 
{thousand}
\cite{amoi6-18},
\cite{amoi6-64})
of copies of that state,
so that we hereafter call this approach 
``multiple-preparation QIP'' 
\ymodifartitionehundredsixvonestepone{(we previously called it ``batch QIP'' in}
\cite{amoi6-104}).
These copies may be
obtained in parallel from an ensemble 
of systems
or successively for the 
{same}
system
(``repeated write/read'', or RWR, procedure
\cite{amoi5-31},
\cite{amoi6-18},
\cite{amoi6-42}).
The above type of measurement is performed for each of these copies
and one counts the number of occurrences of each of the possible
results
$
[
+
\frac{1}{2}
,
+
\frac{1}{2}
,
\dots
,
+
\frac{1}{2}
,
+
\frac{1}{2}
]
$
and so on.
The associated sample relative
frequencies are then used as estimates
of the 
probabilities
\yqubitnbarbspaceoveralleventindexequalstdwithvecbasisprob .
\subsection{%
Single-preparation 
\ymodifartitionehundredsixvonestepone{QIP}%
}
\label{sec-stochastic-QIP-introduce}
The above description was provided for an arbitrarily
selected 
{deterministic pure}
quantum state
\yqubitnbarbtimenonestate .
When developing our first class of BQSS
methods
\cite{amoi5-31},
\cite{amoi6-18},
\cite{amoi6-42},
we had to extend that framework 
to
\emph{random}
{pure quantum}
states.
We especially detailed that concept in
\cite{amoi6-67}.
Briefly,
{the}
coefficients
$
\alpha _i
$
and
$
\beta _i
$
in
(\ref{eq-twoqubit-state-init-index-i-subscript-i})
and
\yqubitnbarbspaceoverallcoefindexequalstdwithvecbasis\
in
(\ref{eq-def-qubitnbarbtimenonestate})
{then}
become complex-valued random variables (RVs).
Hence, 
the
probabilities in
(\ref{eq-def-qubitnbarbspaceoveralleventindexequalstdwithvecbasisprob})
also become 
RVs~!

The problem tackled
in this section is the estimation of some statistical parameters
of these RVs
(\ref{eq-def-qubitnbarbspaceoveralleventindexequalstdwithvecbasisprob}),
namely 
their expectations.
The natural 
(global)
procedure
that may be used to this end,
and that we used in the specific context of BQSS
{\cite{amoi5-31},}
\cite{amoi6-18},
\cite{amoi6-42},
consists of the following two levels.
The lower level only concerns 
{one deterministic}
state
(\ref{eq-def-qubitnbarbtimenonestate})
and the associated
probabilities
(\ref{eq-def-qubitnbarbspaceoveralleventindexequalstdwithvecbasisprob})
{which}
are estimated from a large number of copies
of the considered state, using the 
multiple-preparation
QIP framework 
of
Section
\ref{sec-stochastic-QIP-standard-concepts}.
This is repeated for different states 
(\ref{eq-def-qubitnbarbtimenonestate})
and then, at the higher level,
the sample mean over all these states is separately computed for
each probability
\yqubitnbarbspaceoveralleventindexequalstdwithvecbasisprob\
{%
(with samples
supposedly drawn from the same 
{statistical}
distribution).}
{We here}
aim at proceeding further:
at the above%
{-defined}
lower level, we aim at using \emph{a small number}
of copies of the considered state, or ultimately
\emph{a single} 
instance
of that state,
thus developing 
{what}
we 
call ``single-preparation QIP''
\ymodifartitionehundredsixvonestepone{(we
called it ``stochastic QIP'' in}
\cite{amoi6-104}).
At first sight, it might seem that this is not possible, because the
lower level would thus not provide accurate estimates, 
that one could
then confidently gather at the higher level.
However, 
we claim 
and 
show below 
that this 
approach can
be used
{if}
one only aims at estimating some 
{statistical parameters}
of 
the considered quantum states.

We now first 
build the proposed approach by starting from
the frequentist 
view 
of probabilities
(see e.g.
\cite{book-papoulis})
at the above-defined two levels of the considered 
procedure,
that is:
\begin{itemize}
\item At the higher level, where one combines the contributions
associated with
\ytwoqubitseqnb\
states
of the set of
\yqubitnbarb\
qubits.
These states
are
indexed by
$
\Ytwoqubitseqindex 
\in
\{
1
,
\dots
,
\Ytwoqubitseqnb
\}
$
and denoted as
\yqubitnbarbtimenonestateseqindex .
\item At the lower level, 
{which 
concerns
one
deterministic
}
state
\yqubitnbarbtimenonestateseqindex\
and the associated probabilities
\yqubitnbarbspaceoveralleventindexequalstdwithvecbasisprobseqindex\
defined 
{by
(\ref{eq-def-qubitnbarbspaceoveralleventindexequalstdwithvecbasisprob}) but}
with coefficients
\yqubitnbarbspaceoverallcoefindexequalstdwithvecbasisseqindex\
{which depend}
on 
{state}
\yqubitnbarbtimenonestateseqindex .
\end{itemize}
At the 
lower level, each probability 
\yqubitnbarbspaceoveralleventindexequalstdwithvecbasisprobseqindex\
is defined as
\begin{equation}
\Yqubitnbarbspaceoveralleventindexequalstdwithvecbasisprobseqindex\
=
\lim
_{
\Ywritereadonestatenb
\rightarrow
+
\infty
}
\frac
{
\Yqubitnbarbspaceoveralleventindexequalstdwithvecbasisnboccurseqindexwritereadonestatenb
}
{
\Ywritereadonestatenb
}
\end{equation}
{provided this limit exists.}
\yqubitnbarbspaceoveralleventindexequalstdwithvecbasisnboccurseqindexwritereadonestatenb\
is the number of occurrences 
of event
\yqubitnbarbspaceoveralleventindexequalstdwithvecbasis\
for the state
\yqubitnbarbtimenonestateseqindex\
when performing measurements for a set of
\ywritereadonestatenb\
copies
of that state
\yqubitnbarbtimenonestateseqindex .
In practice, one uses only a \emph{finite} number
\ywritereadonestatenb\
of copies
of state
\yqubitnbarbtimenonestateseqindex\
and therefore only accesses the following approximation
of the above probability:
\begin{equation}
\Yqubitnbarbspaceoveralleventindexequalstdwithvecbasisprobseqindexapproxwritereadonestatenb
=
\frac
{
\Yqubitnbarbspaceoveralleventindexequalstdwithvecbasisnboccurseqindexwritereadonestatenb
}
{
\Ywritereadonestatenb
}
.
\label{eq-def-qubitnbarbspaceoveralleventindexequalstdwithvecbasisprobseqindexapproxwritereadonestatenb}
\end{equation}

The higher level of the considered procedure
then addresses
the statistical mean
associated with
samples,
indexed by
\ytwoqubitseqindex ,
of a given quantity,
which is here theoretically
\yqubitnbarbspaceoveralleventindexequalstdwithvecbasisprobseqindex .
In the frequentist approach, this statistical mean is
defined 
{(if the limit exists)}
as
\begin{equation}
\Yqubitnbarbspaceoveralleventindexequalstdwithvecbasisprobexpect
=
\lim
_{
\Ytwoqubitseqnb
\rightarrow
+
\infty
}
\frac
{
\sum
_{
\Ytwoqubitseqindex
=
1
}
^{
\Ytwoqubitseqnb
}
\Yqubitnbarbspaceoveralleventindexequalstdwithvecbasisprobseqindex
}
{
\Ytwoqubitseqnb
}
.
\end{equation}
At the higher level too,
in practice one uses only a \emph{finite} 
number
\ytwoqubitseqnb\
of states
\yqubitnbarbtimenonestateseqindex ,
which 
first
yields
the following approximation
if only 
{performing an}
approximation at the higher level
of the procedure:
\begin{equation}
\Yqubitnbarbspaceoveralleventindexequalstdwithvecbasisprobexpectapproxonetwoqubitseqnb
=
\frac
{
\sum
_{
\Ytwoqubitseqindex
=
1
}
^{
\Ytwoqubitseqnb
}
\Yqubitnbarbspaceoveralleventindexequalstdwithvecbasisprobseqindex
}
{
\Ytwoqubitseqnb
}
.
\end{equation}
The latter expression may then be modified by replacing
its term
\yqubitnbarbspaceoveralleventindexequalstdwithvecbasisprobseqindex\
by its approximation
(\ref{eq-def-qubitnbarbspaceoveralleventindexequalstdwithvecbasisprobseqindexapproxwritereadonestatenb}).
This yields
\begin{equation}
\Yqubitnbarbspaceoveralleventindexequalstdwithvecbasisprobexpectapproxtwotwoqubitseqnb
=
\frac
{
\sum
_{
\Ytwoqubitseqindex
=
1
}
^{
\Ytwoqubitseqnb
}
\Yqubitnbarbspaceoveralleventindexequalstdwithvecbasisnboccurseqindexwritereadonestatenb
}
{
\Ytwoqubitseqnb
\Ywritereadonestatenb
}
.
\label{eq-def-qubitnbarbspaceoveralleventindexequalstdwithvecbasisprobexpectapproxtwotwoqubitseqnb}
\end{equation}
$
\sum
_{
\Ytwoqubitseqindex
=
1
}
^{
\Ytwoqubitseqnb
}
\Yqubitnbarbspaceoveralleventindexequalstdwithvecbasisnboccurseqindexwritereadonestatenb
$
is nothing but the 
number,
hereafter denoted as
\yqubitnbarbspaceoveralleventindexequalstdwithvecbasisnboccurqubitnbarbstatewritereadseqnb
,
of occurrences of event
\yqubitnbarbspaceoveralleventindexequalstdwithvecbasis\
for the
complete
considered set of
$
\Yqubitnbarbstatewritereadseqnb
=
\Ytwoqubitseqnb
\Ywritereadonestatenb
$
measurements.
Therefore,
\yqubitnbarbspaceoveralleventindexequalstdwithvecbasisprobexpectapproxtwotwoqubitseqnb\
is the 
relative frequency of occurrence of that
event over these
$
\Yqubitnbarbstatewritereadseqnb
$
measurements,
or ```trials'', 
using standard probabilistic terms
{\cite{book-papoulis}.}
This quantity (\ref{eq-def-qubitnbarbspaceoveralleventindexequalstdwithvecbasisprobexpectapproxtwotwoqubitseqnb}) may therefore also be expressed
as
\begin{eqnarray}
\Yqubitnbarbspaceoveralleventindexequalstdwithvecbasisprobexpectapproxtwotwoqubitseqnb
&
=
&
\frac
{
\Yqubitnbarbspaceoveralleventindexequalstdwithvecbasisnboccurqubitnbarbstatewritereadseqnb
}
{
\Yqubitnbarbstatewritereadseqnb
}
\label{eq-qubitnbarbspaceoveralleventindexequalstdwithvecbasisprobexpectapproxtwotwoqubitseqnb-number-in-qubitnbarbstatewritereadseqnb}
\\
&
=
&
\frac
{
\sum
_{
\Yqubitnbarbstatewritereadseqindex
=
1
}
^{
\Yqubitnbarbstatewritereadseqnb
}
\Yqubitnbarbspaceoveralleventindexequalstdwithvecbasisindicfuncqubitnbarbstatewritereadseqindex
}
{
\Yqubitnbarbstatewritereadseqnb
}
\label{eq-qubitnbarbspaceoveralleventindexequalstdwithvecbasisprobexpectapproxtwotwoqubitseqnb-indicfunc}
\end{eqnarray}
where
\yqubitnbarbspaceoveralleventindexequalstdwithvecbasisindicfuncqubitnbarbstatewritereadseqindex\
is the value of the indicator function 
of
event
\yqubitnbarbspaceoveralleventindexequalstdwithvecbasis\
for trial
\yqubitnbarbstatewritereadseqindex ,
which takes the value 1 if
\yqubitnbarbspaceoveralleventindexequalstdwithvecbasis\
occurs during that trial, and 0 otherwise.
When using
(\ref{eq-qubitnbarbspaceoveralleventindexequalstdwithvecbasisprobexpectapproxtwotwoqubitseqnb-indicfunc}),
one now considers the
$
\Yqubitnbarbstatewritereadseqnb
=
\Ytwoqubitseqnb
\Ywritereadonestatenb
$
trials as organized as a single series, with trials
indexed by 
\yqubitnbarbstatewritereadseqindex .
One thus
fuses
the above%
{-defined}
two levels of the
procedure into a single one, thus disregarding the fact
that, in this series, each block of 
\ywritereadonestatenb\
consecutive
trials uses the same state
\yqubitnbarbtimenonestateseqindex .
{One may therefore wonder whether
the number
\ywritereadonestatenb\
of used copies of
each state
\yqubitnbarbtimenonestateseqindex\
may be freely decreased, and even set to one,
while possibly keeping the same
total
number 
\yqubitnbarbstatewritereadseqnb\
of trials.}
A 
formal proof
of the relevance 
{of that approach%
}%
,
using Kolmogorov's view of probabilities%
, is provided in 
\cite{amoi6-104}.
Moreover,
\cite{amoi6-104}
thus proves that the proposed estimator 
(\ref{eq-qubitnbarbspaceoveralleventindexequalstdwithvecbasisprobexpectapproxtwotwoqubitseqnb-indicfunc})
of
\yqubitnbarbspaceoveralleventindexequalstdwithvecbasisprobexpect\
is attractive because,
for states independently randomly drawn with the same
distribution and with one 
instance of each
state,
this estimator is asymptotically efficient, 
that is,
when the number 
\yqubitnbarbstatewritereadseqnb\
of trials
tends to infinity:
(i) the mean of this estimator tends to the actual value
\yqubitnbarbspaceoveralleventindexequalstdwithvecbasisprobexpect ,
i.e. this estimator is asymptotically unbiased
(it is even unbiased for a \emph{finite} number of trials),
and
(ii) the variance of this estimator tends to 0.
\section{Test conditions}
\label{sec-appendix-test-conditions}
All tests reported in Section 
\ref{sec-tests}
were performed in the following conditions.
The 
six parameters
\yparamqubitindexstdstateplusmodulus,
\yparamqubitindexstdstateplusphase\
and
\yparamqubitindexstdstateminusphase,
with
$\Yqubitindexstd \in \{ 1 , 2 \}$,
of each initial state
\ymixsyststateinitial\
were randomly drawn with a uniform distribution,
over an interval which depends on the
step of the considered BQPT method, in order to meet
the constraints on the statistics of these parameters
that are imposed by that BQPT method.
The parameters
\yparamqubitonestateminusmodulus\
and
\yparamqubittwostateminusmodulus\
were
then
derived from
(\ref{eq-paramqubitindexstdstateminusmodulus-vs-paramqubitindexstdstateplusmodulus}).
More precisely,
as a first step,
to estimate
the absolute value of
\ytwoqubitresultphaseevolsin\
as explained in the first part of Section
\ref{sec-estim-v},
the
qubit 
parameter values 
\yparamqubitonestateplusmodulus\
and
\yparamqubittwostateplusmodulus\
were
selected within the 
20\%-80\%
sub-range of their
0\%-100\%
allowed range 
defined by
(\ref{eq-qubit-modulus-allowed-range}),
that is,
$
[
0.1
,
0.4
[
$
for
\yparamqubitonestateplusmodulus\
and
$
[
0.6
,
0.9
[
$
for
\yparamqubittwostateplusmodulus ,
as in
\cite{amoi6-42}.
Besides,
\yparamqubitonestateminusphase\
and
\yparamqubittwostateminusphase\
were
drawn
over
$
[
0
,
{2 \pi [}
$
whereas
$
\Yparamqubitonestateplusphase
$
and
\yparamqubittwostateplusphase\
were fixed to 0
(%
\ymodifartitionehundredsixvonestepone{as stated above,}
the parameters which have a physical meaning are
$
\Yparamqubitindexstdstateminusphase
-
\Yparamqubitindexstdstateplusphase
$).
These data are thus such
that
$
E
\{
\sin \Ytwoqubitresultphaseinit
\}
=
0
$%
,
as required by 
this step of
the considered BQPT method.
Then,
as a second step,
to estimate
the sign of
\ytwoqubitresultphaseevolsin\
as explained in the second part of Section
\ref{sec-estim-v},
the same conditions as in the above first step were used for
\yparamqubitindexstdstateplusmodulus,
\yparamqubitindexstdstateplusphase\
and
\yparamqubitindexstdstateminusphase,
with
$\Yqubitindexstd \in \{ 1 , 2 \}$,
except that
\yparamqubitonestateminusphase\
was fixed to 0
and
\yparamqubittwostateminusphase\
was drawn
over
$
[
0
,
\pi [
$.
These data are thus such
that
$
E
\{
\sin \Ytwoqubitresultphaseinit
\}
$
is non-zero and has a known sign (here, it is positive),
as required by 
this step of
the considered BQPT method.
The above two steps were performed with
$
\Ytwoqubitwritereadtimeintervalindexone
=
0.51
$~ns
\ymodifartitionehundredsixvonestepone{\cite{footnote-tests}}%
.
Finally, to estimate
\ytwoqubitsprobadirxxplusplusdiffminusminusphasediffcosgen\
and
\ytwoqubitsprobadirxxplusplusdiffminusminusphasediffsingen,
the method of Section
\ref{sec-estim-w-both} uses measurements along the
$Oz$ and $Ox$ axes,
with
$
\Ytwoqubitwritereadtimeintervalindextwo
=
2
\Ytwoqubitwritereadtimeintervalindexone
$.
For each of the parameters
\yparamqubitindexstdstateplusmodulus,
\yparamqubitindexstdstateplusphase\
and
\yparamqubitindexstdstateminusphase,
with
$\Yqubitindexstd \in \{ 1 , 2 \}$,
we used the same statistics for
measurements along the
$Oz$ and $Ox$ axes.
For the first equation
(\ref{eq-twoqubitsprobaplusplusdirxx-minus-twoqubitsprobaminusminusdirxx-def-and-explicit-mean}),
\yparamqubitonestateplusmodulus\
and
\yparamqubittwostateplusmodulus\
were drawn over
$
[
0.1
,
0.4
[
$
and
\yparamqubitonestateminusphase\
and
\yparamqubittwostateminusphase\
were drawn over
$
[
-
\pi
/
2
,
\pi
/
2
[
$,
whereas
\yparamqubitonestateplusphase\
and
\yparamqubittwostateplusphase\
were fixed to 0.
For the second equation
(\ref{eq-twoqubitsprobaplusplusdirxx-minus-twoqubitsprobaminusminusdirxx-def-and-explicit-mean}),
\yparamqubitonestateplusmodulus\
and
\yparamqubittwostateplusmodulus\
were drawn over
$
[
0.6
,
0.9
[
$,
whereas
\yparamqubitonestateminusphase ,
\yparamqubittwostateminusphase,
\yparamqubitonestateplusphase,
and
\yparamqubittwostateplusphase\
were selected in the same way as for the first 
equation
(\ref{eq-twoqubitsprobaplusplusdirxx-minus-twoqubitsprobaminusminusdirxx-def-and-explicit-mean}).
All above conditions concern the identification phase.
Then, in the computation phase, we used
$
\Ytwoqubitwritereadtimeintervalindexthree
=
2
\Ytwoqubitwritereadtimeintervalindextwo
$,
as explained in Section
\ref{sec-estimate-process}.

Besides,
the value of
matrix
\yopmix\
was set
as follows.
{Conventional 
Electron Spin Resonance
generally operates at 
$X$ or $Q$ 
bands
(around $10$ and $%
35$
GHz
respectively). For electron spins with 
$g=2$, at $35$
GHz,
the resonance field is near $1.25$~%
T.
In the simulations, we used the
values 
$g=2$ 
and
$B=1$~T.
Concerning the exchange coupling, 
we chose 
$J_{z}/k_{B}=1$~%
K
and 
$%
J_{xy}/k_{B}=0.3$~%
K
(see
Appendix E of 
\cite{amoi6-18}
and 
\cite{ferretti-phys-rev-2005})%
.
}

%% file: artiti106v1_main_APS_for_upload.bbl
\begin{thebibliography}{99}
\bibitem{footnote-tests}
\ymodifartitionehundredsixvonestepone{%
%
%
{As a typical example, we aim at setting
\ytwoqubitwritereadtimeintervalindexone\
to
\ytwoqubitwritereadtimeintervalindexthree\
around 1~ns.
However, choosing
$
\Ytwoqubitwritereadtimeintervalindextwo
=
1
$%
~ns
exactly,
and hence
$
\Ytwoqubitwritereadtimeintervalindexone
=
0.5
$%
~ns,
yields a specific case:
in the step of the proposed BQPT method when
\ytwoqubitwritereadtimeintervalindextwo\ is used
(i.e.,
when estimating
\ytwoqubitsprobadirxxplusplusdiffminusminusphasediffcosgen\
and
\ytwoqubitsprobadirxxplusplusdiffminusminusphasediffsingen),
the value of
\ytwoqubitresultphaseevolsin\
is thus
%
%
%
%
very close to 1, 
%
%
%
which is its maximum possible value,
as shown by
(\ref{eq-def-Ytwoqubitresultphaseevol-versionthree}).
To avoid being in such a specific case, 
we slightly shifted
\ytwoqubitwritereadtimeintervalindexone,
and hence
\ytwoqubitwritereadtimeintervalindextwo\
and
\ytwoqubitwritereadtimeintervalindexthree,
by setting
%
%
%
$
\Ytwoqubitwritereadtimeintervalindexone
=
0.51
$~ns.%
%
%
%
}
}
\bibitem{footnote-no-constraint-rone-rtwo}
In this part of our method,
we do not use
(\ref{eq-qubitindexstdstateplusmodulus-sol-both-from-mlsp})
and we therefore do not set the constraint
(\ref{eq-qubit-modulus-allowed-range}).
\end{thebibliography}
